\documentclass[12pt, a4paper]{article}

\usepackage[height=24cm,width=17cm, centering]{geometry}

\usepackage{graphicx}     
\usepackage[bookmarksopen,colorlinks=true,linkcolor=light_blue,
citecolor=light_pink,urlcolor=dark_green,linktocpage=false]{hyperref}

\usepackage{amsfonts}
\usepackage{amsmath}
\usepackage{amssymb}
\usepackage{slashed}
\usepackage{bm,here}
\usepackage{cite}
\usepackage{comment}
\usepackage{cases}
\usepackage{slashed}
\usepackage{colortbl}
\usepackage{xcolor}
\usepackage[utf8]{inputenc}
\usepackage[mathscr]{eucal}
\usepackage[utopia]{mathdesign}

\usepackage{multicol}
\definecolor{dark_blue}{rgb}{0.0, 0., 0.6}
\definecolor{dark_red}{rgb}{0.7, 0., 0.}
\definecolor{dark_green}{rgb}{0., 0.45, 0.3}
\definecolor{light_pink}{rgb}{1,0.4,0.4}
\definecolor{light_blue}{rgb}{0.284602,0.317763,0.963947}
\definecolor{red}{rgb}{1,0,0}
\definecolor{blue}{rgb}{0,0,1}
\definecolor{orange}{rgb}{1,0.5,0}

\begin{document}


\newcommand{\vev}[1]{ \left\langle {#1} \right\rangle }
\newcommand{\bra}[1]{ \langle {#1} | }
\newcommand{\ket}[1]{ | {#1} \rangle }
\newcommand{\eV}{ \ {\rm eV} }
\newcommand{\KeV}{ \ {\rm keV} }
\newcommand{\MeV}{\  {\rm MeV} }
\newcommand{\GeV}{\  {\rm GeV} }
\newcommand{\TeV}{\  {\rm TeV} }
\newcommand{\1}{\mbox{1}\hspace{-0.25em}\mbox{l}}
\newcommand{\Red}[1]{{\color{red} {#1}}}

\newcommand{\lmk}{\left(}  
\newcommand{\rmk}{\right)}
\newcommand{\lkk}{\left[}  
\newcommand{\rkk}{\right]}
\newcommand{\lhk}{\left \{ }  
\newcommand{\rhk}{\right \} }
\newcommand{\del}{\partial}  
\newcommand{\la}{\left\langle} 
\newcommand{\ra}{\right\rangle}
\newcommand{\half}{\frac{1}{2}}

\newcommand{\dd}{\mathrm{d}}
\newcommand{\Mpl}{M_{\rm Pl}}
\newcommand{\mg}{m_{3/2}}
\newcommand{\abs}[1]{\left\vert {#1} \right\vert}
\newcommand{\mphi}{m_{\phi}}
\newcommand{\Hz}{\ {\rm Hz}}
\newcommand{\for}{\quad \text{for }}
\newcommand{\Min}{\text{Min}}
\newcommand{\Max}{\text{Max}}
\newcommand{\Kahler}{K\"{a}hler }
\newcommand{\cphi}{\varphi}
\newcommand{\Tr}{\text{Tr}}
\newcommand{\diag}{{\rm diag}}

\newcommand{\SUf}{SU(3)_{\rm f}}
\newcommand{\Upq}{U(1)_{\rm PQ}}
\newcommand{\Zpq}{Z^{\rm PQ}_3}
\newcommand{\Cpq}{C_{\rm PQ}}
\newcommand{\ubar}{u^c}
\newcommand{\dbar}{d^c}
\newcommand{\ebar}{e^c}
\newcommand{\nubar}{\nu^c}
\newcommand{\Ndw}{N_{\rm DW}}
\newcommand{\Fpq}{F_{\rm PQ}}
\newcommand{\fpq}{v_{\rm PQ}}
\newcommand{\Br}{{\rm Br}}
\newcommand{\Lag}{\mathcal{L}}
\newcommand{\Lqcd}{\Lambda_{\rm QCD}}
\newcommand{\const}{\text{const}}

\newcommand{\ji}{j_{\rm inf}} 
\newcommand{\jb}{j_{B-L}} 
\newcommand{\M}{M} 
\newcommand{\im}{{\rm Im} }
\newcommand{\re}{{\rm Re} }
\newcommand{\cm}{\ {\rm cm} }

\def\lrf#1#2{ \left(\frac{#1}{#2}\right)}
\def\lrfp#1#2#3{ \left(\frac{#1}{#2} \right)^{#3}}
\def\lrp#1#2{\left( #1 \right)^{#2}}
\def\REF#1{Ref.~\cite{#1}}
\def\SEC#1{Sec.~\ref{#1}}
\def\FIG#1{Fig.~\ref{#1}}
\def\EQ#1{Eq.~(\ref{#1})}
\def\EQS#1{Eqs.~(\ref{#1})}
\def\blue#1{\textcolor{blue}{#1}}
\def\red#1{\textcolor{blue}{#1}}

\newcommand{\fa}{f_{a}}
\newcommand{\Uh}{U(1)$_{\rm H}$}
\newcommand{\osc}{_{\rm osc}}

\newcommand{\mav}{\left. m_a^2 \right\vert_{T=0}}
\newcommand{\mat}{m_{a, {\rm QCD}}^2 (T)}
\newcommand{\mam}{m_{a, {\rm M}}^2 }
\def\eq#1{Eq.~(\ref{#1})}
\def\fig#1{Fig.~\ref{#1}}

\newcommand{\LQCD}{\Lambda_{\rm QCD}}

\newcommand{\UH}{U(1)$_H$ }

\newcommand{\EV}{ \ {\rm eV} }
\newcommand{\KEV}{ \ {\rm keV} }
\newcommand{\MEV}{\  {\rm MeV} }
\newcommand{\GEV}{\  {\rm GeV} }
\newcommand{\TEV}{\  {\rm TeV} }

\def\order#1{\mathcal{O}(#1)}

\makeatletter
\@addtoreset{equation}{section}
\def\theequation{\thesection.\arabic{equation}}
\makeatother

\newcommand{\TODO}[1]{{\color{red}{$[[ \clubsuit\clubsuit$ \bf #1 $\clubsuit\clubsuit ]]$}}}
\newcommand{\my}[1]{\textcolor{blue}{[{\bf MY}: #1]}}

\begin{titlepage}

\begin{center}

\hfill IPMU 17-0069\\

\vskip 1.in

{\Huge \bf 
False Vacuum Decay\\[.5em]
Catalyzed by Black Holes
}

\vskip .55in

{\large Kyohei Mukaida$^\blacklozenge$ and Masaki Yamada$^\lozenge$}

\vskip .35in

\begin{tabular}{ll}
$^{\blacklozenge}$ &\!\! {\em Kavli IPMU (WPI), UTIAS,}\\
&{\em The University of Tokyo,  Kashiwa, Chiba 277-8583, Japan}\\[.3em]
$^{\lozenge}$ &\!\! {\em Institute of Cosmology, Department of Physics and Astronomy,}\\
&\!{\em Tufts University, Medford, MA  02155, USA}\\[.3em]

\end{tabular}

\end{center}
\vskip .5in

\begin{abstract}
\noindent
False vacuum states are metastable in quantum field theories, and true vacuum bubbles can be nucleated due to the quantum tunneling effect. It was recently suggested that an evaporating black hole (BH) can be a catalyst of bubble nucleations and dramatically shortens the lifetime of the false vacuum. In particular, in the context of the Standard Model valid up to a certain energy scale, even a single evaporating BH may spoil the successful cosmology by inducing the decay of our electroweak vacuum. In this paper, we reinterpret catalyzed vacuum decay by BHs, using an effective action for a thin-wall bubble around a BH to clarify the meaning of bounce solutions. We calculate bounce solutions in the limit of a flat spacetime and in the limit of negligible backreaction to the metric, where it is much easier to understand the physical meaning, and compare these results with the full calculations done in the literature. As a result, we give a physical interpretation of the enhancement factor: it is nothing but the probability of producing states with a finite energy. This makes it clear that all the other states such as plasma should also be generated through the same mechanism, and calls for finite-density corrections to the tunneling rate which tend to stabilize the false vacuum. We also clarify that the dominant process is always consistent with the periodicity indicated by the BH Hawking temperature after summing over all possible remnant BH masses or bubble energies, although the periodicity of each bounce solution as a function of a remnant BH can be completely different from the inverse temperature of the system as mentioned in the previous literature.
\end{abstract}

\end{titlepage}

\tableofcontents
\thispagestyle{empty}
\renewcommand{\thepage}{\arabic{page}}
\renewcommand{\thefootnote}{$\natural$\arabic{footnote}}
\setcounter{footnote}{0}

\newpage
\setcounter{page}{1}

\section{Introduction and Summary
\label{sec:introduction}}

The discovery of the Higgs boson has established the Standard Model (SM)~\cite{Aad:2012tfa, Chatrchyan:2012xdj}. For the current center value of the SM parameters, especially the measured top and Higgs masses~\cite{Giardino:2013bma, Aad:2013wqa, Chatrchyan:2013mxa, Aad:2015zhl, Khachatryan:2015hba}, the Higgs potential develops a lower energy state than the electroweak vacuum at around the intermediate scale well below the Planck~\cite{Sher:1988mj, Anderson:1990aa, Arnold:1989cb, Arnold:1991cv, Espinosa:1995se, Isidori:2001bm, Espinosa:2007qp, Ellis:2009tp, EliasMiro:2011aa, Bezrukov:2012sa, Degrassi:2012ry, Buttazzo:2013uya, Bednyakov:2015sca}. This fact implies that the quantum tunneling might lead to a disastrous decay of our vacuum~\cite{Coleman:1977py, Coleman:1980aw}. Fortunately, it is known that, for the current favored value of SM parameters, its lifetime far exceeds the present age of the Universe, and thus our electroweak vacuum is believed to be metastable in the context of SM valid up to a very high energy scale.

Though this argument guarantees the safety of our vacuum in the present Universe, it does not mean that our metastable vacuum can survive throughout the history of the Universe. Thus, this scenario could in principle contradict various cosmological phenomena which can drive the vacuum decay. Many studies have been performed from this viewpoint. For instance, in the early Universe, it is believed that the Universe was filled with thermal plasma composed of SM particles. Since the Higgs interacts with SM particles including the Higgs itself, the thermal fluctuations might activate the decay of our vacuum~\cite{Linde:1981zj, Sher:1988mj, Arnold:1991cv, Espinosa:1995se} while these relativistic particles tend to stabilize the Higgs at the same time. It has been shown that this effect does not spoil our Universe for the best fit values of SM parameters~\cite{Espinosa:2007qp, Ellis:2009tp, EliasMiro:2011aa}. If we further go back through the history of the Universe, we may encounter the phase of inflation and the subsequent (p)reheating. Since light fields with masses smaller than the Hubble parameter acquire fluctuations proportional to the Hawking temperature $H / 2\pi$ during inflation~\cite{Guth:1982ec, Starobinsky:1982ee, Hawking:1982cz, Bardeen:1983qw, Hawking:1982my, Starobinsky:1994bd}, the Higgs might overcome the potential barrier via the Hawking-Moss instanton~\cite{Hawking:1981fz}, which may be interpreted as the thermal hopping owing to the Hawking temperature. This observation puts a severe bound on the Hubble parameter of inflation: $H < \mathcal O (0.01) \Lambda_\text{inst}$ with $\Lambda_\text{inst}$ being the instability scale~\cite{Espinosa:2007qp, Lebedev:2012sy, Enqvist:2013kaa, Kobakhidze:2013tn, Herranen:2014cua, Hook:2014uia, Kobakhidze:2014xda, Kamada:2014ufa, Enqvist:2014bua, Fairbairn:2014zia, Espinosa:2015qea, Kearney:2015vba, Kohri:2016qqv, East:2016anr, Joti:2017fwe}. Although one might think that this constraint can be ameliorated by introducing a small coupling between the inflaton and the Higgs field, recent studies reveal that the (p)reheating stage after inflation could drive the catastrophic decay because the very interaction activates Higgs fluctuations due to the oscillating inflaton~\cite{Herranen:2015ima, Ema:2016kpf, Kohri:2016wof, Enqvist:2016mqj, Postma:2017hbk, Ema:2017loe}.

Recently, it was pointed out that a black hole (BH) can be a nucleation site just like a boiling stone in a superheated liquid system, and the vacuum transition rate can be dramatically enhanced (or the potential barrier becomes effectively smaller) around the BH~\cite{Gregory:2013hja, Burda:2015isa, Burda:2015yfa} (see Refs.~\cite{Hiscock:1987hn, Berezin:1987ea, Arnold:1989cq, Berezin:1990qs, Gomberoff:2003zh, Garriga:2004nm} for earlier work). 
The result of their calculation is independent of the periodicity of the Wick-rotated time coordinate, 
so that they insist that the result can be applied to an arbitrary low temperature system. 
In particular, the enhancement gets more significant for a smaller BH, and in an extreme case, the Higgs can classically overcome the potential barrier, such as the thermal hopping and the Hawking-Moss transition. Applying this result to the Higgs field, they concluded that even a single small BH that evaporates within the current age of the Universe leads to the disaster of our vacuum~\cite{Burda:2016mou}.\footnote{
	See also Refs.~\cite{Tetradis:2016vqb, Canko:2017ebb} for related work. 
}
And thus, there should not be such a small BH in our observable Universe. Although such a small BH may not be formed in the usual scenario of cosmology, their conclusion puts a stringent constraint on some cases, such as the formation of primordial BHs in the early Universe~\cite{Gorbunov:2017fhq}.\footnote{
	The bubble nucleation process is also important in the context of the multiverse, 
	where bubbles continuously nucleate and observers may live in the baby universes~\cite{Sato:1981bf, Maeda:1981gw, Sato:1981gv, Berezin:1982ur, Blau:1986cw, Garriga:1997ef, Garriga:2015fdk, Deng:2016vzb}. 
	The enhancement effect of the nucleation rate is applied in Ref.~\cite{Oshita:2016btk} 
	to generate baby universes around BHs. 
}

In this paper, we reinterpret the earlier results derived via a Wick-rotated Euclidean field theory in Refs.~\cite{Gregory:2013hja, Burda:2015isa, Burda:2015yfa} by invoking an effective action for a thin-wall bubble that can also describe the vacuum transition in scalar field theories~\cite{Berezin:1987ea, Arnold:1989cq, Basu:1991ig}. 
We start with the bubble nucleation at a finite temperature in the flat spacetime, and recall that the final bubble nucleation rate can be factorized into the probability of producing states with a finite energy times a tunneling rate of a finite energy.
This is also true if we have a BH.
By extending the same procedure,
we reformulate the bubble nucleation rate around a BH in the case where the backreaction of the bubble on the spacetime can be neglected.\footnote{
	This situation is practically important for realistic applications,
	for instance, to study the metastable Higgs vacuum~\cite{Burda:2015isa}.
}

We compare the bubble nucleation rate computed in this way 
with the full gravitational one in the limit of negligible backreaction.
As a result, we clarify the meaning of the enhancement factor, that is, a probability of producing states with a finite energy $E$, which is a Boltzmann factor $e^{- E /T_{\rm BH}}$.
It is hard to imagine that a BH only activates bubbles since quantum field theory has many other degrees of freedom to be excited.
Hence, we expect all the states with a finite energy $E$ should also be generated by the same mechanism.
This argument clarifies the need of finite-density corrections to the bubble nucleation rate regardless of its origin, namely whether or not the Universe is filled by the plasma of the BH Hawking temperature,
though the size of corrections depends on it.\footnote{
	See also Sec.~\ref{conclusions}.
}

We also confirm that the periodicity of each bounce solution as a function of $E$ is not necessarily related to the temperature of the system.
However, after summing over all the possible transitions as a function of $E$,
we find that the dominant process is always consistent with the periodicity indicated by the temperature of the system.
This observation also holds if we have a BH.
Although, one still cannot determine the question raised in Ref.~\cite{Gorbunov:2017fhq}, whether or not the thermal plasma fills the whole Universe, by only looking at the periodicity of bounce solutions,
our procedure indicates that the heart of the problem is free from a BH.
The problem is whether or not a finite-volume heat reservoir can emit bubbles 
whose size is much larger than the size of the reservoir.
We leave this issue as an open question.

The following is the summary of our results of this paper: 
\begin{itemize}
\item In a flat spacetime with a finite-temperature plasma, we have shown $- \dd B/ \dd E_* = 2 \Delta \tau$ and $\dd^2 B / \dd E_*^2 < 0$, where $-B$ is the exponent of the quantum tunneling rate, $E_*$ is the energy of the bubble, and $2 \Delta \tau$ is the periodicity of the bubble solution. Since the exponent of the Boltzmann factor satisfies $\dd (-E_* / T_*) / \dd E_* = -1/T_*$, the inequality $\dd^2 B / \dd E_*^2 < 0$ implies that the dominant process is given either by $E_* = 0$ or $E_* = E_{\rm sp}$, where $E_{\rm sp}$ is the sphaleron energy. 

\item In the Schwarzschild--de Sitter spacetime, we eventually find $- \dd B_{\rm bubble} / \dd \Delta M = 2 \Delta \tau$ and $-\dd B_{\rm boundary} / \dd \Delta M = -1/T_{\rm BH, -}$, where $T_{\rm BH, -}$ is the Hawking temperature associated with the remnant BH and $-B_{\rm boundary}$ and $-B_{\rm bubble}$ are the exponents of the quantum tunneling rate coming from the boundary of BH and the other contributions, respectively. The difference of the BH mass before and after the transition, $\Delta M$, is equal to the bubble energy by the conservation of energy. We also show that $B_{\rm bubble} (\Delta M)$ coincides with $B (E_* = \Delta M)$ in the limit where the bubble radius is much larger than the BH radius once we identify the temperature as the Hawking temperature. In particular, $\dd^2 B_{\rm bubble} / \dd \Delta M^2 < 0$ in that limit. 

\item In the fixed-background Schwarzschild--de Sitter spacetime with finite-temperature effects, we again obtain $- \dd B/ \dd E_* = 2 \Delta \tau$. The behavior of the second derivative is similar to the above full calculation. In the case that the effect of the change of the metric by the bubble is negligible, 
the nucleation rate coincides with the one derived by the above full calculation only if we identify the temperature of the system as the Hawking temperature of the BH.
This observation clarifies that the enhancement factor is nothing but the probability of generating states with a finite energy, which is the Boltzmann factor with a BH Hawking temperature.
\end{itemize}

This paper is organized as follows. In Sec.~\ref{without gravity}, we first review the calculation of the tunneling rate for a thin-wall bubble in a scalar field theory. 
We show that the transition rate is dominated either by a vacuum transition without an excited energy or by a sphaleron transition in this system. Next, we take into account gravity and consider the vacuum transition in the Schwarzschild-de Sitter spacetime in Sec.~\ref{BH}. In particular, we calculate the bubble energy dependence of transition rate and show that its behavior is similar to the one in a finite-temperature system in a flat spacetime. We also use the effective action for the thin-wall bubble and show that the same nucleation rate in the literature can be derived by the thermal activation of the BH Hawking temperature in a certain limit. Section~\ref{conclusions} is devoted to the conclusion and discussion, 
where 
we briefly explain the physics behind our result and discuss the possibility that the cost of such thermal plasma may significantly reduce the bubble nucleation rate.

\section{Transition without gravity
\label{without gravity}}

In this section, we review the calculation of the transition rate from a false vacuum to a true vacuum in quantum field theory without gravity, \textit{i.e.}, in the limit of $G \to 0$, where $G$ $[\equiv 1/( 8 \pi \Mpl^2) ]$ is the Newton constant  and $\Mpl$ ($= 2.4 \times 10^{18} \GeV)$ is the Planck scale. We take gravity into account in Sec.~\ref{BH}.

\subsection{Tunneling from a false vacuum}

The action is given by 
\begin{align}
 S \lkk \phi \rkk = \int \dd^4 x \lkk - \frac{1}{2} \del_\mu \phi \del^\mu \phi - V(\phi) \rkk,
\end{align}
where $V(\phi)$ is a potential for the scalar field, which has a false vacuum at $\phi = \phi_{\rm FV}$ 
and the true vacuum at $\phi = \phi_{\rm TV}$.

The lifetime of the false vacuum 
can be calculated from the path integral as follows: 
\begin{align}
 e^{- \Gamma t_0} 
 &=
 \frac{\abs{\la \phi_{\rm bubble}, t= t_0 \vert \phi_{\rm FV}, t= 0 \ra}^2}
 {\abs{\la \phi_{\rm FV}, t=t_0 \vert \phi_{\rm FV}, t=0 \ra}^2}
 \\
  &= 
 \abs{ \int_{\phi (t=0) = \phi_{\rm FV}}^{\phi (t= t_0) = \phi_{\rm bubble}} \mathcal{D} \phi e^{iS [\phi]} }^2 
 \Bigg/ 
  \abs{ \int_{\phi (t= 0) = \phi_{\rm FV}}^{\phi (t= t_0) = \phi_{\rm FV}} \mathcal{D} \phi e^{iS [\phi]} }^2 
  \label{no bounce}
 \\
 &= 
 \int_{\rm bounce} \mathcal{D} \phi e^{iS [\phi]} 
 \Bigg/ 
 \int_{\phi (t \in [- t_0, t_0]) = \phi_{\rm FV}} \mathcal{D} \phi e^{iS [\phi]} 
 \\ 
 &=
 \int_{\rm bounce} \mathcal{D} \phi e^{i S [\phi] - i S_{M,0}}, 
 \label{Gamma for QFT} 
 \\[.5em]
 S_{\rm M,0} &\equiv S [\phi (x) = \phi_{\rm FV}], 
\end{align}
with $t_0 \to \infty$, 
where the path integral $\int_{\rm bounce}$ 
is performed under the boundary conditions of $\phi (t = \pm t_0) = \phi_{\rm FV}$ and $\phi(t = 0) = \phi_{\rm bounce}$. 
The subscript ``bubble" in $\phi$ means that 
it is a bubble configuration with a certain radius as we specify below. 
The denominator in the first line comes from the normalization of the initial and final states 
and gives the factor $e^{- i S_{\rm M, 0}}$ in the last line, 
where the subscript ``M'' indicates this action is defined in Minkowski spacetime.

Now we take the imaginary time $\tau = i t$ 
and rewrite \eq{Gamma for QFT} in terms of the Euclidean path integral. 
In the saddle point approximation in Euclidean theory, 
the path integral is approximated by $S_E [\phi_{\rm bounce}]$, 
which is calculated from a bounce solution to the classical Euclidean equation of motion. 
The action is minimized by a solution where $\phi$ bounces only once. 
In addition to the single bounce solution, 
there is an infinite number of solutions where $\phi$ bounces many times, 
which may be summed in the dilute gas approximation. 
Then we obtain 
\begin{align}
e^{- \Gamma t_0} 
&\propto
 e^{- \abs{K} t_0 \exp{- (S_E[\phi_{\rm bounce}] - S_E[\phi = \phi_{\rm FV}])}}, 
\end{align}
where $\abs{K}$ is a prefactor that is not important for our discussion. 
Thus we obtain 
\begin{align}
 \Gamma &\propto e^{-B},
 \\
 B &= S_\text{bounce} - S_{\text{E},0}, 
\end{align}
where 
the Euclidean action is given by 
\begin{align}
 S_E [\phi] = \int \dd^4 x \lkk \frac{1}{2} \del_\mu \phi \del_\mu \phi + V(\phi) \rkk,
\end{align}
and the normalization factor is given by 
\begin{align}
 S_{\text{E},0} = S_E [ \phi = \phi_{\rm FV}]. 
\end{align}

Let us emphasize that the action $S_\text{bounce}$ is calculated from the bounce solution under the boundary condition of $\phi = \phi_{\rm FV}$ at $\tau = \pm \tau_0 (\to \pm \infty)$, and the tunneling process corresponds to a transition from $\phi = \phi_{\rm FV}$ to $\phi = \phi_{\rm bounce}$, \textit{i.e.},  a transition from the metastable ground state $\phi= \phi_{\rm FV}$ to the state at the other side of the potential barrier $\phi = \phi_{\rm bounce}$. 
Although we mention here the contribution from the perturbation $\delta \phi (\tau)$, 
it does not usually contribute to the exponential factor.\footnote{
	If the system couples with light degrees of freedom, the prefactors could be significant~\cite{Caldeira:1982uj,PhysRevB.29.130,Weinberg:1992ds,Matsumoto:2010tg,Garbrecht:2015yza}.
	In other words, we have to be careful what the ``tree-level'' action is in computing the bounce.
	Throughout this paper, we do not consider this issue further, 
	and simply assume that we somehow know the tree-level action that is
	appropriate to compute the bounce.
}

The bounce solution $\phi_{\rm bounce}$ obeys 
the Euclidean equation of motion that is given by the variational principle of $S_E [\phi]$ in terms of $\phi$: 
\begin{align}
 &\frac{\dd^2 \phi}{\dd \tau^2} + \Delta \phi + U ' = 0, 
 \\
 &U(\phi) = - V(\phi), 
\end{align}
where $\Delta = \del_i^2$ is the Laplacian. 
In quantum field theory, 
the degrees of freedom is infinity because of the spacial dependence of the field. 
In many cases, however, 
we can use some symmetries to reduce the degrees of freedom 
to unity.

The Euclidean action has an O(4) symmetry 
in quantum field theory, 
so let us first focus on O(4) symmetric solutions. 
In the thin-wall approximation,  the scalar field configuration is approximated by the following O(4) symmetric 
configuration: 
\begin{align}
 \phi (x) = 
 \phi_{\rm thin} (\eta; \eta_*) \equiv 
 \left\{ 
 \begin{array}{ll}
 \phi_{\rm TV}
 &\quad \text{for} \ \eta \ll \eta_*, 
 \\
 \phi_{\rm FV} 
 &\quad \text{for} \ \eta \gg \eta_*, 
 \end{array}
 \right.
 \label{phi(r)}
\end{align}
where $\eta$ ($\equiv \sqrt{\tau^2 + x^2 + y^2 + z^2}$) is the radial coordinate of the Euclidean spacetime
and $\eta_*$ is the radius of the bubble. 
Note that this is an instanton solution 
where the time variable $\tau$ runs from $- \infty$ to $\infty$ 
and the configuration is nontrivial in a small interval $\abs{\tau} \lesssim \eta_*$.

The number of degrees of freedom is reduced to be unity by the O(4) symmetry, 
so that we can consider a one-dimensional system with the variable $\eta_*$. 
Plugging the O(4) symmetric thin-wall configuration back into the action,
we can express the action as a function of the bubble radius $\eta_\ast$, 
\begin{align}
S_E= - \frac{1}{2} \pi^2 \eta_*^4 \epsilon + 2 \pi^2 \eta_*^3 \sigma + S_{\text{E},0}, 
\end{align}
where the first term comes from the contribution inside the bubble, 
while the second term comes from the surface of the bubble. 
We define the energy density difference as $\epsilon \equiv V (\phi_{\rm FV}) - V (\phi_{\rm TV})$
and the surface energy density of bubble $\sigma$ as
\begin{align}
 \sigma \equiv \int_{\phi_{\rm TV}}^{\phi_{\rm FV}} \dd \phi \sqrt{2 \lkk V (\phi) - V(\phi_{\rm FV}) \rkk}. 
 \label{sigma}
\end{align}
The variable $\eta_*$ obeys a constraint that originates from the Euclidean equation of motion.
The same equation can be derived from the variational principle of $S_E[ \phi( \eta; \eta_*) ]$ in terms of the variable $\eta_*$. 
Then we find the following results: 
\begin{align}
 \eta_* &= \eta_0 \equiv \frac{3 \sigma}{\epsilon},
 \label{eta_0}
 \\
 S_E &= \frac{27 \pi^2}{2} \frac{\sigma^4}{\epsilon^3} + S_{\text{E},0} (\phi_\text{FV}).
\end{align}
We find 
\begin{align}
 B = B_0 \equiv \frac{27 \pi^2}{2} \frac{\sigma^4}{\epsilon^3}. 
 \label{B_0}
\end{align}

It is instructive to consider the same theory with an O(3) spherical symmetric assumption, 
which can be generalized to calculate transition rates in a finite temperature. 
In the thin-wall approximation, 
the worldsheet metric is written as (see \textit{e.g.}, Ref.~\cite{Basu:1991ig}) 
\begin{align}
 \dd s^2_3 = - \lkk 1 - \lmk \frac{\dd r_*}{\dd t} \rmk^2 \rkk \dd t^2 
 + r_*^2 \dd \Omega. 
 \label{worldsheet metric}
\end{align}
The action for the domain wall is given by the 
worldsheet area in addition to the difference of potential energy inside and outside bubble, 
so that we obtain 
\begin{align}
 S = \int \dd t \lkk - 4 \pi r_*^2 \sigma \gamma^{-1} + \frac{4}{3} \pi r_*^3 \epsilon \rkk + S_{M,0}, 
 \label{S_E QFT}
\end{align}
where $\gamma = \lkk 1 - (\dd r_* / \dd t )^2\rkk^{- 1/2}$ 
may be regarded as the gamma factor of the domain wall. 
The first term is the surface term, which contains the kinetic energy of bubble, 
and the second term comes from the contribution inside the bubble. 
The conserved energy can be derived from 
\begin{align}
 E = \frac{\del L }{\del \dot{r}_*} \dot{r}_* - L, 
\end{align}
where the Lagrangian $L$ can be read from the above action. 
This is rewritten as 
\begin{align}
 4 \pi r_*^2 \sigma \gamma - \frac{4}{3} \pi r_*^3 \epsilon &= E_\ast \label{finite_energy} \\ 
 &= 0.
\end{align}
In the second equality, we use the fact that the initial energy is zero. 
Taking the imaginary time $\tau = i t$, 
this can be rewritten as 
\begin{align}
 \lmk \frac{\dd r_*}{\dd \tau} \rmk^2 = \lmk \frac{3 \sigma}{\epsilon} \rmk^2 \frac{1}{r_*^2} - 1. 
\end{align}
The bounce solution is given by 
\begin{align}
 r_* = \lkk \lmk 3 \sigma / \epsilon \rmk^2 - \tau^2 \rkk^{1/2}, 
\end{align}
so that we obtain the value of the Euclidean bounce action as 
\begin{align}
 S_E = \frac{27\pi^2}{2} \frac{\sigma^4}{\epsilon^3} + S_{\text{E},0}. 
\end{align}
Noting that $r_*^2 + \tau^2 = ( 3 \sigma / \epsilon)^2 = \eta_*^2$, 
these results are consistent with the above results using the O(4) symmetric assumption.

\subsection{Tunneling with a finite energy \label{sec2-3}}

\begin{figure}[t]
\centering 
\includegraphics[width=.45\textwidth]{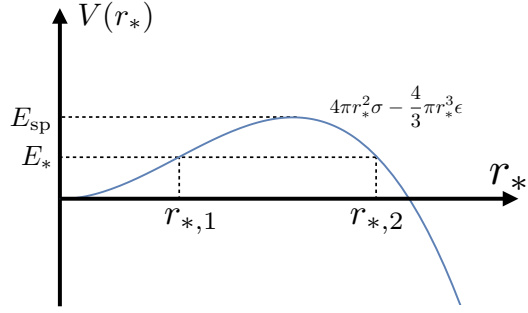} 
\caption{\small
 Example of bubble potential as a function of its radius $r_\ast$.
}
  \label{fig:3}
\end{figure}

Now we can consider a transition from an excited state around the false vacuum 
by using the O(3) approximation and the thin-wall approximation. 
As one can see from Eq.~\eqref{finite_energy} and Fig.~\ref{fig:3},
a state with a finite energy $E_\ast$ allows an O(3) symmetric bubble with $\mathrm d r / \mathrm d t = 0$,
whose radius is $r_{\ast 1}$ or $r_{\ast,2}$.
The amplitude of the transition from a bubble with $r_* = r_{*,1}$ at $\tau = 0$ 
to another one with $r_* = r_{*2}$ at $\tau = \tau_0$, which is going to expand,
is obtained from 
\begin{align}
 \la r_{*,2} , \tau = \tau_0 \vert r_{*,1}, \tau = 0 \ra 
 = 
 \int_{r (\tau = 0) = r_{*,1}}^{r (\tau = \tau_0) = r_{*,2}} \mathcal{D} \lkk \phi \rkk e^{-S [\phi]}.
\end{align}
The transition rate is then given by 
\begin{align}
 e^{- \Gamma t_0} &\propto 
  \abs{ \la r_{*,2}, \tau = \tau_0 \vert r_{*,1} \tau = 0 \ra }^2 
  \\
&  =
 \abs{ \int_{r (\tau = 0) = r_{*,1}}^{r (\tau = \tau_0) = r_{*,2}} \mathcal{D} \lkk \phi \rkk e^{-S [\phi]} }^2 \\ 
& =
 \int_{\rm bounce} \mathcal{D} \lkk \phi \rkk e^{-S [\phi]}, 
\end{align}
where the last path integral is performed under the boundary conditions such that 
$r = r_{*,1}$ at $\tau = -\tau_0$, 
$r = r_{*,2}$ at $\tau = 0$, 
and $r = r_{*,1}$ at $\tau = \tau_0$. 
By using the saddle-point approximation, 
the path integral is replaced by the dominant contribution $e^{- S[\phi_{\rm bounce}]}$ 
where $\phi_{\rm bounce}$ is the bounce solution obeying the Euclidean equation of motion 
with the boundary condition of $r_* \in [ r_{*,1} , r_{*2}]$. 
The amplitude is normalized by 
$\left\vert \la r_{*,1}, \tau = \tau_0 \vert r_{*,1}, \tau = 0 \ra \right\vert^2$.
As a result, the transition rate can be expressed as
\begin{align}
 \Gamma  
 \propto 
 e^{-B (E_*)}, \quad B (E_*) = S_\text{bounce} (E_\ast) - S_{\text{E},0} (E_\ast), 
\end{align}
where $S_{\text{E},0} (E_\ast)$ comes from the normalization. 
Note again that $S_\text{bounce} (E_\ast)$ is obtained from the Euclidean equation of motion
from $r_{\ast, 1}$ to $r_{\ast, 2}$ while $S_{\text{E},0} (E_\ast)$ is an O(3) symmetric bubble 
with a fixed radius $r_{\ast, 1}$.

The bounce with a finite energy $E_\ast$ satisfies 
\begin{align}
 4 \pi r_*^2 \sigma \gamma - \frac{4}{3} \pi r_*^3 \epsilon = E_*, 
 \label{energy conservation law2} 
\end{align}
where $E_*$ is the initial energy.
There are two solutions with a vanishing wall velocity $\dot r = 0$, $r_{\ast,1}$ and $r_{\ast,2}$,
which are obtained from $E_\ast = 4 \pi \sigma r_{\ast}^2 - 4 \pi r_{\ast}^3 \epsilon /3$ as can be seen from Fig.~\ref{fig:3}.
Note here that the initial energy should be small enough to have these two solutions.
The critical energy, above which we do not have solutions to Eq.~\eqref{finite_energy},
is obtained from the condition $\dd r_{*} / \dd \tau = \dd^2 r_{*} / \dd \tau^2 = 0$.
The critical solution is
\begin{align}
 r_{*, {\rm sp}} &= \frac{2 \sigma}{\epsilon},
 \\
 E_{\rm sp} &= \frac{16 \pi \sigma^3}{3\epsilon^2}.
\end{align}
Here, the subscript ``sp'' indicates that this is nothing but the sphaleron, as we will see in the next Sec.~\ref{sec:tunneling_thermal}.
When we regard $r_*$ as a position variable of a particle in a one-dimensional system, 
the constraint \eq{energy conservation law2} can be rewritten as 
the following conservation law of ``energy": 
\begin{align}
 &\frac{1}{2} \lmk \frac{\dd r_*}{\dd \tau} \rmk^2 + U (r_*) = 0, 
 \\
 &2 U  (r_*) = 1 - \lkk \frac{\epsilon}{3 \sigma} r_* + \frac{E_*}{4\pi r_*^2 \sigma} \rkk^{-2}. 
\label{U(r) in QFT}
\end{align}
Plugging the solutions into Eq.~\eqref{S_E QFT}, we get the bubble nucleation rate for $E_\ast < E_{\text{sp}}$, 
\begin{align}
 B (E_*) 
 &= \int \dd r_* \sqrt{\lmk 4 \pi r_*^2 \sigma \rmk^2 - \lmk \frac{4}{3} \pi r_*^3 \epsilon + E_* \rmk^2} 
 \\
 &=
 \int \dd \tau 4 \pi r_*^2 \sigma \gamma \lmk \frac{\dd r_*}{\dd \tau} \rmk^2, 
\end{align}
where we use $S_{\text{E},0} (E_\ast) = E_* \int \dd \tau + S_{\text{E}, 0}$. 
Once we regard the factor $4 \pi r_*^2 \sigma \gamma$ as the effective mass of the bubble, 
the result is similar to that in the one-dimensional quantum mechanical system.

Note again that the above transition means that
a bubble with a radius $r_{*,1}$, which is not $r_{*}= 0$, tunnels into the one with a radius $r_{*,2}$.
Hence, we need to specify the way to excite the initial state to 
the bubble with the radius $r_* = r_{*,1}$. 
If such bubbles are continuously produced and collapse in the initial state 
with a finite probability, the vacuum decay rate may be expressed as 
the probability of creating bubbles with $r_* = r_{*,1}$
times the probability of tunneling from $r_* = r_{*,1}$ to $r_* = r_{*,2}$, namely $e^{-B (E)}$. 
Here, note that we first assume the field configuration as \eq{phi(r)} 
and reduce the number of degrees of freedom to unity. 
Since there are infinite degrees of freedom for the scalar field, 
it is generally difficult to give the energy so that all the energy is converted to such a macroscopic configuration, 
that is, the initial bubble with a radius $r_{\ast,1}$.
The thermal state is an example that we have such an excited initial condition naturally. 
In this case, all degrees of freedom have a typical energy of order $T_*$ with the Boltzmann weight, 
and hence the probability of creating the initial bubble is nothing but $\exp ( - E_\ast (r_* = r_{*,1}) /T)$ and is nonzero.

\subsection{Tunneling with a thermal energy}
\label{sec:tunneling_thermal}

Now we can consider 
a transition in the scalar field theory in a thermal background with a temperature of $T_*$. 
The transition rate can be calculated 
by the integral of the Boltzmann factor times quantum tunneling rate (see \textit{e.g.}, Ref.~\cite{Brown:2007sd}): 
\begin{align}
 \Gamma &= \Gamma_q + \Gamma_c, 
 \label{gamma_q and c}
 \\
 \Gamma_q &\sim \int_{0}^{E_\text{sp}} \dd E 
 e^{- E/ T_*} e^{-B (E)},  
 \label{Boltzmann factor}
  \\
 \Gamma_c &\sim
 \int_{E_\text{sp}}^\infty \dd E 
 e^{-E / T_*}, 
 \label{classical_qft}
\end{align}
where $\Gamma_c$ is the classical transition rate.
Note that, for $E_\ast > E_\text{sp}$, the bubble nucleation rate is unity, $B(E) = 1$,
where $E_\text{sp}$ is the sphaleron energy as explained below. 

Let us evaluate the transition rate approximately.
As discussed in the case of one-dimensional quantum mechanics,
the question traces back to the behavior of $\dd^2 B / \dd E^2$.
If $\dd^2 B / \dd E^2 > 0$ for $0 \leq E \leq E_\text{sp}$, one may evaluate the integral via the steepest descent method
by expanding the exponent as follows:
\begin{align}
	\frac{E}{T_\ast} + B (E) = \frac{E_\text{cr}}{T_\ast} + B (E_\text{cr}) 
	+ \left[ \frac{1}{T_\ast} + B'(E_\text{cr}) \right] \left( E_\ast - E_\text{cr} \right)
	+ \frac{B'' (E_\text{cr})}{2} \left( E - E_\text{cr} \right)^2 + \cdots.
\end{align}
If one finds the solution to 
\begin{align}
	1/T_\ast 
	&= - B'(E_\text{cr}) 
	\\
	 &=  2\int_{r_* (\tau = -\Delta \tau) = r_{*,1}}^{r_* (\tau = 0) = r_{*,2}} \dd r_* \lmk \abs{ \frac{\dd r_*}{\dd \tau} } \rmk^{-1}
 \\	
	&= 2 \Delta \tau \quad \text{for}~~0 \leq E \leq E_\text{sp},  
	\label{T delta tau}
\end{align}
where $2\Delta \tau$ is the time periodicity of the bounce solution, 
the integral may be approximated by the Gaussian.
In this case the first integral is dominated by the energy $E_\text{cr}$:
\begin{align}
 \Gamma_q
 &\sim
 e^{- E_{\rm cr} /T_*} e^{- (S_{\rm bounce} (E_{\rm cr}) - S_{\text{E},0}(E_{\rm cr}))} 
 \label{two parts1}
 \\
 &=
 e^{- S_{\rm bounce} (E_{\rm cr}) }, 
 \label{Gamma in thermal system}
\end{align}
where we use $S_{\text{E},0} (E_{\rm cr}) = E_{\rm cr}/T_*$. 
This equation indicates that
the transition rate can be decomposed into two parts as \eq{two parts1}: 
the Boltzmann factor and the quantum tunneling rate. 
On the one hand, the quantum tunneling rate tells one that the bubble with $r_* = r_{*,1}$, which is not equal to $r_* = 0$, tunnels to $r_* = r_{*,2}$. 
On the other hand, the Boltzmann factor represents the probability 
that the bubble with energy $E_{\rm cr}$ reaches the point of $r_* = r_{*,1}$. 
Therefore, in total, 
\eq{Gamma in thermal system} 
gives the transition rate where the bubble 
goes from $r_* = 0$ to $r_* = r_{*,1}$ via thermal excitation 
and then to $r_* = r_{*,2}$ via quantum tunneling. 
If there is no solution to \eq{T delta tau}, the integral is dominated by the boundary between Eq.~\eqref{Boltzmann factor} and Eq.~\eqref{classical_qft},
which ends up $e^{- E_\text{sp}/T_\ast}$.

\begin{figure}[t]
\centering 
\includegraphics[width=.45\textwidth]{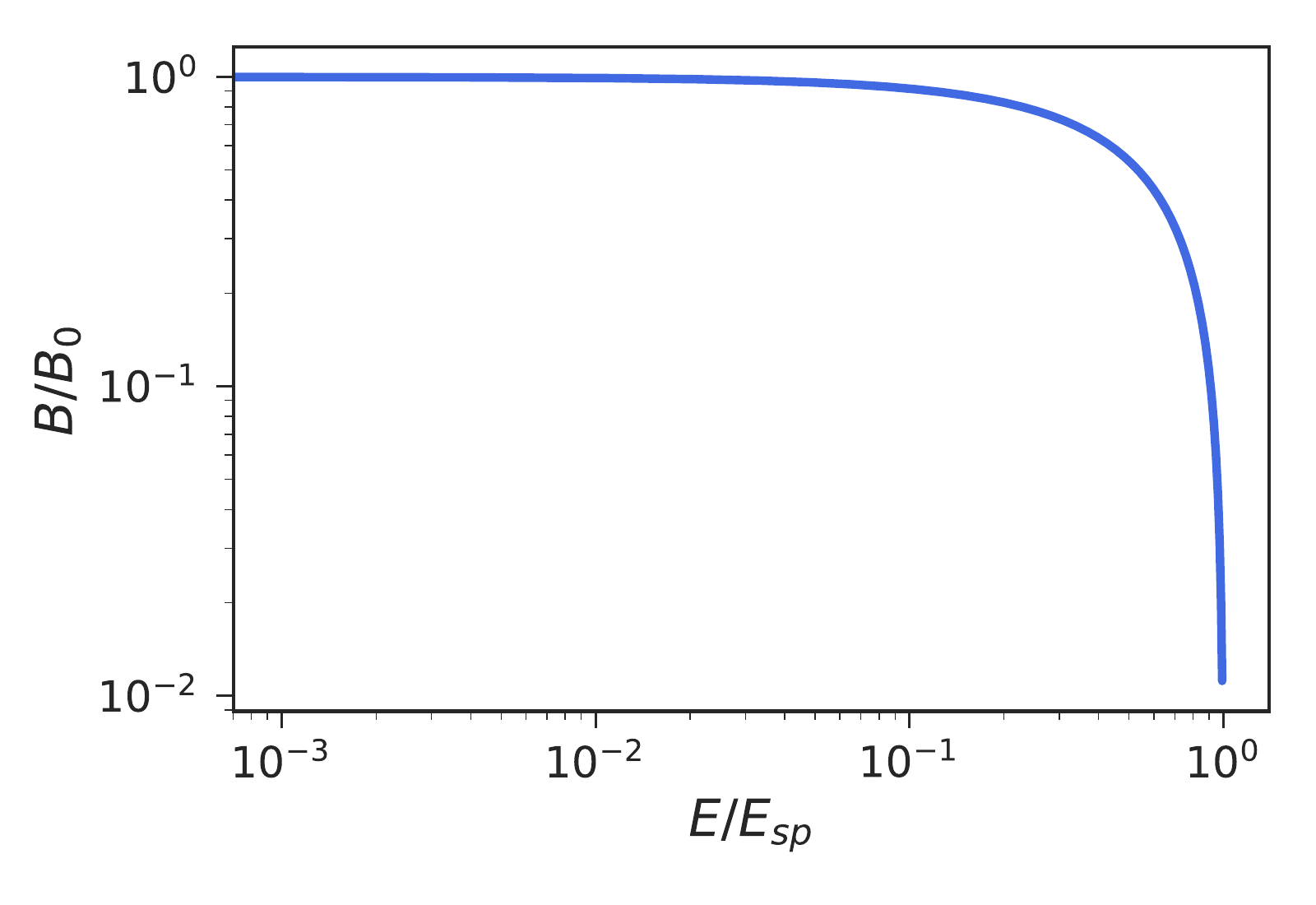} 
\qquad 
\includegraphics[width=.45\textwidth]{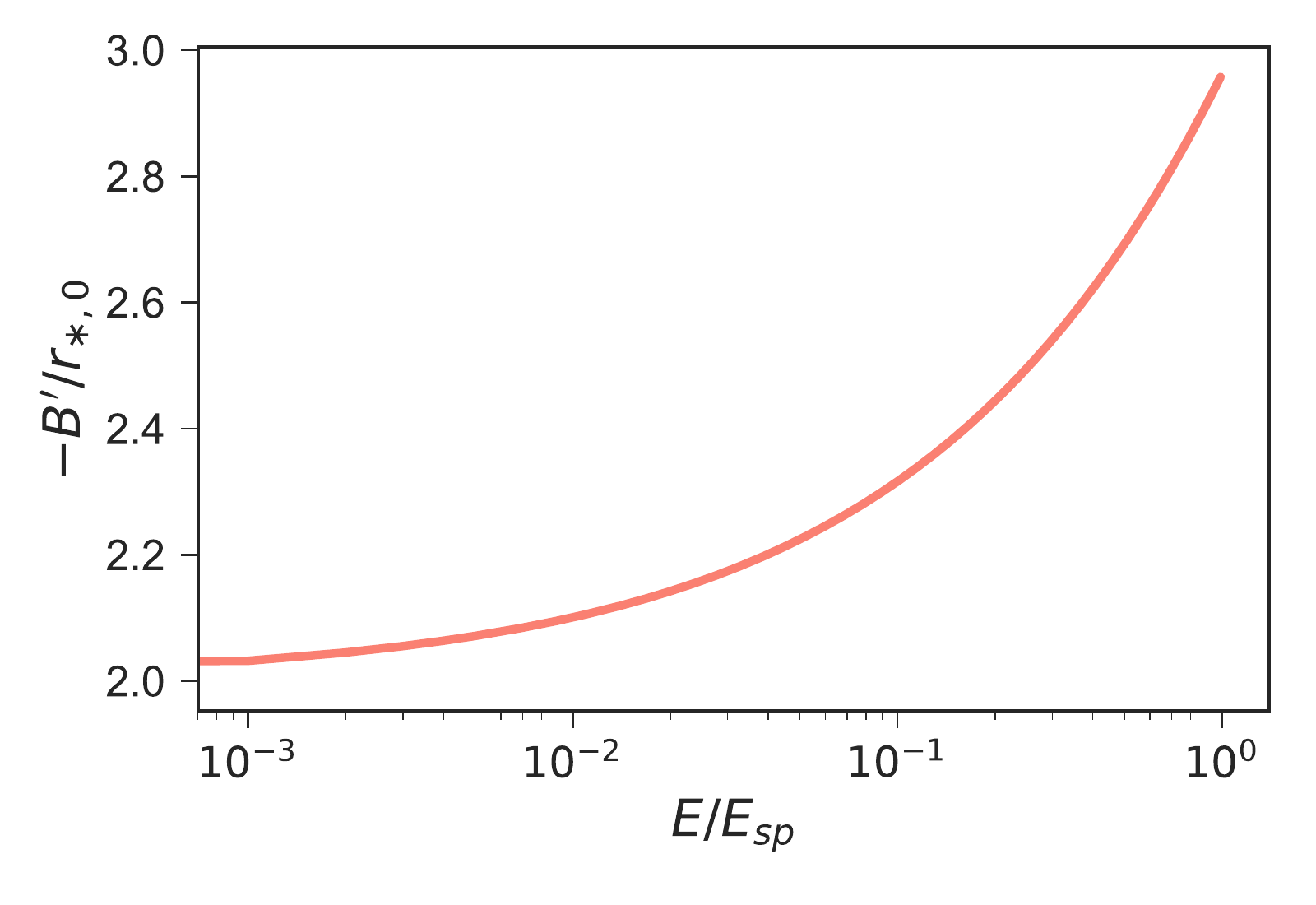} 
\caption{\small
 Plot of $B(E)$ (blue line) and $-\dd B(E) / \dd E$ (pink line) as a function of $E$ 
 for a thin-wall bubble 
 in quantum field theory. 
 Here, we have normalized quantities by those at the zero-temperature:
 $B_0 = 27 \pi^2 \sigma^4 / 2 \epsilon^3$ and $r_{\ast,0} = 3 \sigma / \epsilon$. 
 The results with these combinations are independent of $\sigma$ and $\epsilon$. 
}
  \label{fig:B_E}
\end{figure}

However, at least for the thermal transition in quantum field theory under the thin-wall approximation
with the O$(3)$ symmetry,
the second derivative of the bounce action with respect to its energy
is always $\dd^2 B / \dd E^2 \leq 0$.
Let us first confirm this property.
Figure~\ref{fig:B_E} shows the bounce $B$ and its first derivative, $- B'(E)$, as a function of the energy $E$.
One can see that $- B'(E)$ is an increasing function with respect to $E$,
and thus $\dd B^2 / \dd E^2 \leq 0$.
Therefore, the saddle point is not a minimum of the exponent in Eq.~\eqref{Boltzmann factor}.
Rather, the integral is dominated by its edges, $E = 0$ or $E_\text{sp}$.
As a result, the transition rate can be expressed as
\begin{align}
	\Gamma &\sim \Gamma_q + \Gamma_c, \\
	\Gamma_q &\sim e^{- B(0)},\\
	\Gamma_c &\sim e^{- E_\text{sp}/T_\ast}.
\end{align}
Recalling the sphaleron energy $E_\text{sp} = 16 \pi \sigma^3/3 \epsilon^2$
and the bounce action $B(0) = 27\pi^2 \sigma^4 / 2 \epsilon^3$, 
we can go further.
It is clear that the sphaleron transition dominates the decay process when 
the temperature $T_*$ is large enough to satisfy $E_{\rm sp} / T_* < B(0)$. 
The threshold temperature is given by 
\begin{align}
	T_{*, \rm th} 
	&= \frac{E_{\rm sp}}{B(0)} 
	= \frac{32 \epsilon}{81 \pi \sigma}\\ 
	&= \frac{32}{27 \pi} \eta_\ast^{-1} \ll \eta_\ast^{-1} .
	\label{T_th}
\end{align}
Thus we conclude that the transition rate is summarized as 
\begin{numcases}
	{\Gamma \sim}
		e^{-B(0)} & for $T_\ast \leq T_{\ast, \text{th}}$, \label{thermal_quantum} \\
		e^{- E_\text{sp}/T_\ast} & for  $T_{\ast, \text{th}} \leq T_\ast$. \label{thermal_classical}
\end{numcases}

Before closing this section,
we would like to explain the relation between the above results and the well-known method of  
putting the theory on $S_{T_\ast^{-1}}^1 \times \mathbb R^3$ (\textit{i.e.}, periodic Wick-rotated time $\tau \sim \tau + T_\ast^{-1}$) and
evaluating the imaginary part of the free energy $F = \ln Z$~\cite{Linde:1981zj}. 
In this case, the geometry forces all the configurations to be periodic, including the bounce.
The quantum one, Eq.~\eqref{thermal_quantum}, corresponds to the dominant periodic instanton.
Interestingly,
though there exist other branches of periodic instantons with a finite energy $E$,
the vacuum one, $E=0$, dominates for the thin-wall approximation under the O$(3)$ symmetry.
Note here that,
since $T_{\ast, \text{th}}^{-1}$ is much larger than the radius of the vacuum bubble $\eta_\ast$
as can be seen from Eq.~\eqref{T_th},
the vacuum bubble may be embedded in $S_{T_\ast^{-1}}^1 \times \mathbb R^3$
for $T_\ast \leq T_{\ast,\text{th}}$.
The classical one, Eq.~\eqref{thermal_classical}, is obtained by the dimensional reduction of $S^1_{T_{\ast}^{-1}}$,
which is a good approximation if the energy scale of the bubble is much smaller than the temperature.
It is clear that the static solution can be embedded in the periodic spacetime.

\section{Transition in the Schwarzschild--de Sitter spacetime 
\label{BH}}

Gravity changes the spacetime in accordance with finite-energy objects.
In the case of our interest, not only the BH but the bubble could be the origin of such distortions.
Roughly speaking, in the context of the vacuum decay,
the change of the spacetime affects the cost to nucleate the bubble.
One can guess that gravity should modify the vacuum decay rate.
Therefore, in this section, we switch on gravity and investigate its effect on the vacuum decay.
The action is given by~\cite{Hawking:1995fd}
\begin{align}
 S 
 &= S_{\rm bubble} + S_G,
 \\
 S_{\rm bubble} &=
 \int \dd^4 x \sqrt{-g} \lkk - \frac{1}{2} \lmk \del \phi \rmk^2 - V(\phi) \rkk ,
 \\
 S_G 
 &= \frac{1}{2} \int_Y R \sqrt{-g} \dd^4 x 
 + S_{\rm boundary},
 \label{S_G}
\end{align}
where $g$ is the determinant of the metric and $R$ is the Ricci scalar. 
We take the Planck unit,
$\Mpl$ ($\simeq 2.4 \times 10^{18} \GEV$) $=1$ (\textit{i.e.}, 
the Newton constant $G$ is taken to be $8 \pi G = 1$), unless otherwise stated.
The curvature scalar $R$ contains terms with second derivatives, 
which can be removed by integration by parts 
and write the action only by first derivatives 
so that we can use the path integral approach in the gravitational theory. 
As a result, the boundary term $S_{\rm boundary}$ arises, 
which is the integral of the trace of the second fundamental form of the boundary $K$~\cite{Gibbons:1976ue}: 
\begin{align}
 S_{\rm boundary} = 
 \int_{\del Y} K \sqrt{-h} \dd^3 x = \frac{\del}{\del n} \int_{\del Y} \sqrt{-h} \dd^3 x 
 = \frac{\del}{\del n} V_{\rm boundary}, 
\end{align}
where $h$ is the determinant of the three-dimensional metric on the boundary surface, 
$V_{\rm boundary}$ is the volume of the boundary, 
and $n$ is the unit normal. 
Note that the region of boundary depends on the metric, 
and hence the boundary term $S_{\rm boundary}$ is determined only after we specify the metric.

In particular,
we consider a bubble nucleation in (anti-)de Sitter spacetime with a BH, 
which is described by the Schwarzschild-de Sitter metric, 
\begin{align}
 \dd s^2 &= - f_{\rm SdS} (r) \dd t^2 + \frac{\dd r^2}{f_{\rm SdS}(r)} + r^2 \dd \Omega, 
 \\
 f_{\rm SdS}(r) &= 1 - \frac{M_{\rm BH}}{4 \pi r} - \frac{\Lambda r^2}{3}, 
 \label{SdS metric}
\end{align}
where $M_{\rm BH}$ is a BH mass and $\Lambda$ is a vacuum energy. 
The areas of the boundary, $\del V_{\rm boundary}/ \del n$,
at the surface of the BH ($\equiv A_{\rm BH}$) 
and at the cosmological horizon ($\equiv A_{\rm dS}$) 
are given by $A_{\rm BH} \simeq M_{\rm BH}^2 / 4 \pi$ and 
$A_{\rm dS} \simeq 4 \pi / H^2$ in this metric, respectively, where 
$H$ is the inverse of the apparent horizon length related to the vacuum energy 
as $H^2 = \Lambda / 3$. 
This metric respects only an O(3) symmetry.
We use the O(3) symmetric assumption to find bounce solutions in the following analysis.

\subsection{Bubble nucleation via tunneling \label{sec3-1}}

Here we illustrate how to evaluate the bounce action in the presence of a BH with
gravitational backreaction following Refs.~\cite{Gregory:2013hja, Burda:2015isa, Burda:2015yfa, Burda:2016mou}. 
Since we employ a thin-wall approximation,
we may use the Euclidean metric defined separately in the outer and inner regions of the bubble: 
\begin{align}
 \dd s^2 =& f_\pm (r) \dd \tau_\pm^2 + \frac{\dd r^2}{f_\pm(r)} + r^2 \dd \Omega,
 \\
 f_\pm(r) =& 1 - \frac{M_{\pm}}{4 \pi r} - \frac{\Lambda_\pm r^2}{3}, 
\end{align}
where $M_{+}$ is the initial BH mass and 
$M_{-}$ is the remnant BH mass after the bubble nucleation. 
The zeros of $f_\pm$ define the horizon of each patch.
We have at most two zeros for each $f_\pm$: the BH horizon $R_{\text{BH},\pm}$ 
and the de Sitter horizon $R_{\text{dS},\pm}$ for $\Lambda_{\pm} > 0$
respectively.
The natural periodicities to eliminate the conical deficits at each horizon are the following:
$T_{\rm BH,\pm}^{-1} = 4 \pi R_{\rm BH,\pm} / (1 - \Lambda_\pm R_{\rm BH,\pm}^2)$ at $r = R_{\rm BH, \pm}$
and $T_{\rm dS,\pm}^{-1} = 4 \pi R_{\rm dS,\pm} / (\Lambda_\pm R_{\rm dS,\pm}^2 - 1)$ at $r = R_{\rm dS, \pm}$.

We now specify the setup of our interest.
In the following, we focus on the case where
we initially have a BH with $M_+$. 
Also, we assume that the vacuum energy of the scalar field changes from $\Lambda_+$ to $\Lambda_-$ due to this transition.
On the other hand, the remnant BH mass $M_-$ is taken to be an arbitrary parameter. 
We do not impose that the Hawking temperature of the initial black hole 
$T_{\text{BH},+} = 4 \pi R_{\rm BH, +}/(1 - \Lambda_+ R_{\rm BH, +}^2)$ 
should coincide with that inferred from the periodicity of the bubble solution.
This mismatch will cause a conical deficit at the BH horizon, and thus we have to cope with it appropriately
as done in Ref.~\cite{Gregory:2013hja}.

Before evaluating the bounce action,
we briefly illustrate how to obtain the wall trajectory of the nucleated bubble
with fixed $M_+$, $\Lambda_\pm$, and $T_\text{ini}$.
The wall trajectory $r_* (\lambda)$ is parametrized by the proper time of 
a comoving observer of the wall,
\begin{align}
 f_\pm \dot{\tau}_\pm^2 + \frac{\dot{r}_*^2}{f_\pm} = 1, 
\end{align}
where the dot denotes the derivative with respect to the proper time $\lambda$. 
The Israel junction condition yields
\begin{align}
 f_+ \gamma_+ - f_- \gamma_- = - \frac{1}{2} \sigma r_*, 
\end{align}
where $\sigma$ is the tension of the bubble 
and 
\begin{align}
 \gamma_\pm \equiv \dot{\tau}_\pm = 
 \frac{1}{\sqrt{f_\pm(r_*) + \frac{1}{f_\pm(r_*)} \lmk \frac{\dd r_*}{\dd \tau_\pm} \rmk^2}}. 
\end{align}
We can explicitly rewrite it as 
\begin{align}
 f_\pm \gamma_\pm = \lmk \frac{\Delta \Lambda}{3 \sigma} \mp \frac{\sigma}{4} \rmk r 
 + \frac{\Delta M}{4 \pi \sigma r^2}. 
\end{align}

The junction condition implies that 
the wall velocity has to satisfy the following conservation law of ``energy'': 
\begin{align}
& \frac{1}{2} \lmk \frac{\dd \tilde{r}_*}{\dd \tilde{\lambda}} \rmk^2 + U (\tilde{r}) = 0, 
 \label{EOM with BH}
\\
 & 2 U(\tilde{r}) = 
 \lmk \tilde{r}_* + \frac{k_2}{\tilde{r}_*^2} \rmk^2 
 + \frac{k_1}{\tilde{r}_*} - 1, 
 \label{eq:EOM_potential_BH}
\end{align}
where $\tilde{r}_* = \alpha r_* / \gamma$ and $\tilde{\lambda} = \alpha \lambda / \gamma$, and 
\begin{align}
 k_1&= \frac{\alpha M_-}{4 \pi \gamma} + \frac{(1 - \alpha) \alpha \Delta M}{2 \pi \sigma \gamma^2}, \quad
 k_2 = \frac{\alpha^2 \Delta M}{4 \pi \sigma \gamma^2},
 \quad
 \gamma_{\rm GMW} = \frac{\sigma l^2}{1 + \sigma^2 l^2 / 4} ,
 \quad
 \alpha^2 = 1 + \frac{\Lambda_- \gamma^2}{3},
 \quad
 l^2 = \frac{3}{\Delta \Lambda}, 
\end{align}
where $\Delta M \equiv M_+ - M_-$ and $\Delta \Lambda \equiv \Lambda_+ - \Lambda_-$ ($\equiv \epsilon$).\footnote{
	Note that the $\gamma_{\rm GMW}$ here is different from the gamma factor of the domain wall, $\gamma$, defined below. 
}
Once we fix the all the parameters $M_\pm$ and $\Lambda_\pm$,
we can obtain the wall trajectory as a function of the proper time, $r_\ast (\lambda)$, in principle.
In our setup, we have fixed $M_+$ and $\Lambda_\pm$
and hence we have a family of solutions as a function of the remnant BH mass, $r_\ast (\lambda; M_-)$.

Now we are in a position to evaluate the Euclidean action by the solution to \eq{EOM with BH}.
The gravitational Euclidean action is given by 
\begin{align}
 S_G = S_{\rm boundary} 
 - \frac{1}{2} \lkk \int_{Y_+} \sqrt{g} R + \int_{Y_-} \sqrt{g} R \rkk 
 + \lkk \int_{\del Y_+} \sqrt{h} K + \int_{\del Y_-} \sqrt{h} K \rkk, 
\end{align}
where $Y_-$ and $Y_+$ represent the regions inside and outside of the bubble, respectively, 
and $\del Y_\pm$ represents the boundary induced by the bubble. 
Note again that $S_\text{boundary}$ accounts for the boundaries at the horizons.
The Einstein equation and 
the Israel junction condition imply that 
the action can be rewritten as 
\begin{align}
 S_G = S_{\rm boundary} 
  - \frac{1}{2} \lkk \int_{Y_+} \sqrt{g} 4 \Lambda_+ + \int_{Y_-} \sqrt{g} 4 \Lambda_- \rkk 
 - \frac{3}{2} \lkk \int_{\del Y} \sqrt{h} \sigma \rkk. 
 \label{S_G full}
\end{align}
In the thin-wall approximation, 
the bubble Euclidean action is given by 
\begin{align}
 S_{\rm bubble} 
 &= 
  \int \dd^4 x \sqrt{g} \lkk \frac{1}{2} \lmk \del \phi \rmk^2 + V(\phi) \rkk 
  \\
 &=
 \int \dd \lambda 
 \lkk 4 \pi r_*^2 \sigma 
 + \frac{4 \pi}{3} r_*^3 \lmk \Lambda_- \gamma_- - \Lambda_+ \gamma_+ \rmk 
 - \frac{4 \pi}{3} R_{\rm BH, -}^3 \Lambda_- \gamma_- 
 + \frac{4 \pi}{3} R_{\rm dS, +}^3 \Lambda_+ \gamma_+ 
 \rkk. 
\end{align}
If there is no de Sitter horizon, we have to drop the last term.

The bubble nucleation rate around the BH is calculated from 
\begin{align}
 B = S_\text{E} - S_{\text{E},0}, 
\end{align}
where 
$S_\text{E}$ is the total Euclidean action $(= S_G + S_{\rm bubble})$ 
and 
$S_{\text{E},0}$ is the action without the bubble. 
If the gravitational action changes, the boundary term may also change. 
Assuming that the bubble is not as large as the de Sitter horizon,
one can see that the boundary from the BH horizon only contributes to the difference.
The explicit form of $B$ is given by
\begin{align}
 &B (M_-) 
 = B_{\rm bubble} (M_-) + B_{\rm boundary} (M_-),
 \label{nucleation rate of GMW}
 \\
 &B_{\rm bubble} (M_-) = 
4 \pi \int r_* \lmk f_+ \dd \tau_+  - f_- \dd \tau_- \rmk 
 + \int \frac{4\pi}{3} r_*^3 \lmk \Lambda_+ \dd \tau_+ - \Lambda_- \dd \tau_- \rmk 
 - \frac{1}{2} \int \lmk M_+ \dd \tau_+ - M_- \dd \tau_- \rmk ,
 \label{eq:bubble_BH}
 \\
 &B_{\rm boundary} (M_-) = 
  8 \pi^2 \lmk R_{\rm BH, +}^2 - R_{\rm BH, -}^2 \rmk, 
\end{align}
where we use the Israel junction $\sigma r_* / 2 = - ( f_+ \dot{\tau}_+  - f_- \dot{\tau}_-)$. 
The third term in $B_{\rm bubble}$ is related to the contribution of the conical deficit. 
The term $B_{\rm boundary}$ accounts for the change of the BH entropy which comes from the area of the boundary at the horizon.
We regard this result as a function of $M_-$ for later convenience.
In summary, the bubble nucleation rate from the initial state, 
(a BH of $M_+$,
vacuum energy of $\Lambda_+$),
to the final state, (a BH of $M_-$, vacuum energy $\Lambda_-$),
is given by
\begin{align}
	\Gamma (M_-) \sim e^{-B (M_-)}. 
\end{align}

We emphasize that 
the initial mass of the BH $M_+$, the initial and final energy densities $\Lambda_\pm$, 
and the surface energy density of the bubble $\sigma$ 
are determined by the initial conditions and the potential of the scalar field, 
while the remnant BH mass $M_-$ after the bubble nucleation is not yet determined in the above calculation. 
In the sense of the path integral approach, 
we should sum over all the nucleation rates $\Gamma (M_-) \sim e^{-B(M_-)}$ in terms of the variable $M_-$. 
Thus, 
we need to find a minimal value of the action in terms of $M_-$, 
\begin{align}
 \Gamma \sim \int \dd M_- \Gamma (M_-) 
 \sim \int \dd M_- e^{- B_\text{boundary} - B_\text{bubble}}
 \sim e^{-B_{\rm min}}.
 \label{gamma with BH}
\end{align}
As has been pointed out in Ref.~\cite{Gregory:2013hja}, 
it is possible that the remnant BH mass $M_-$ is larger than the initial BH mass $M_+$ 
though it usually gives subdominant contributions. 
Here we comment on the lower bound on the integral \eq{gamma with BH}. 
If $M_+$ is sufficiently large, 
there is a lower bound of $M_-$, below which 
the ``potential" $U(\tilde{r})$ is always larger than zero 
and 
there is no solution to \eq{EOM with BH}. 
At the critical point, $U (\tilde{r}) = U'(\tilde{r}) = 0$ 
and the solution is static. 
See also the discussion below.

\begin{figure}[t]
\centering 
\includegraphics[width=.45\textwidth]{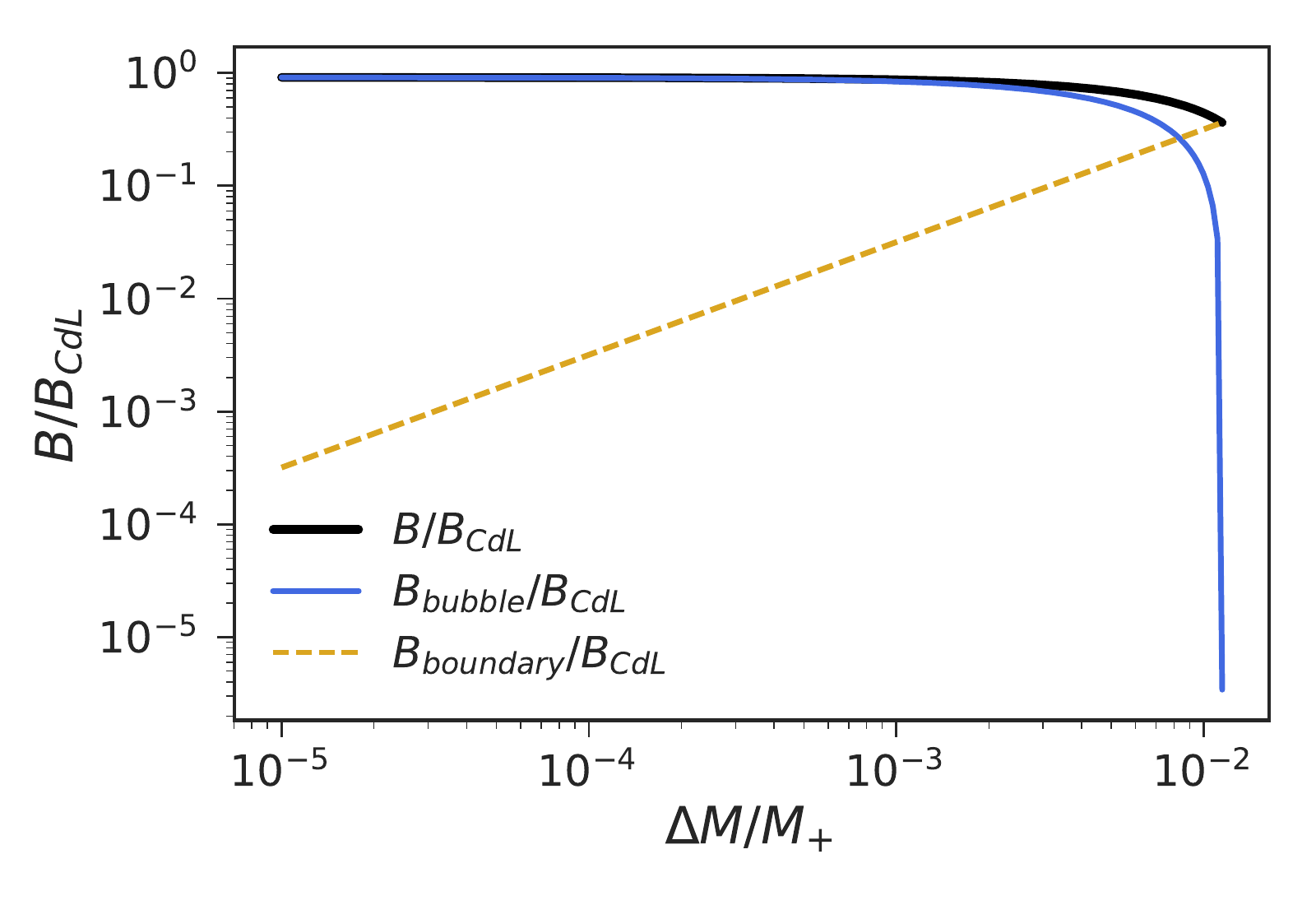} 
\qquad 
\includegraphics[width=.45\textwidth]{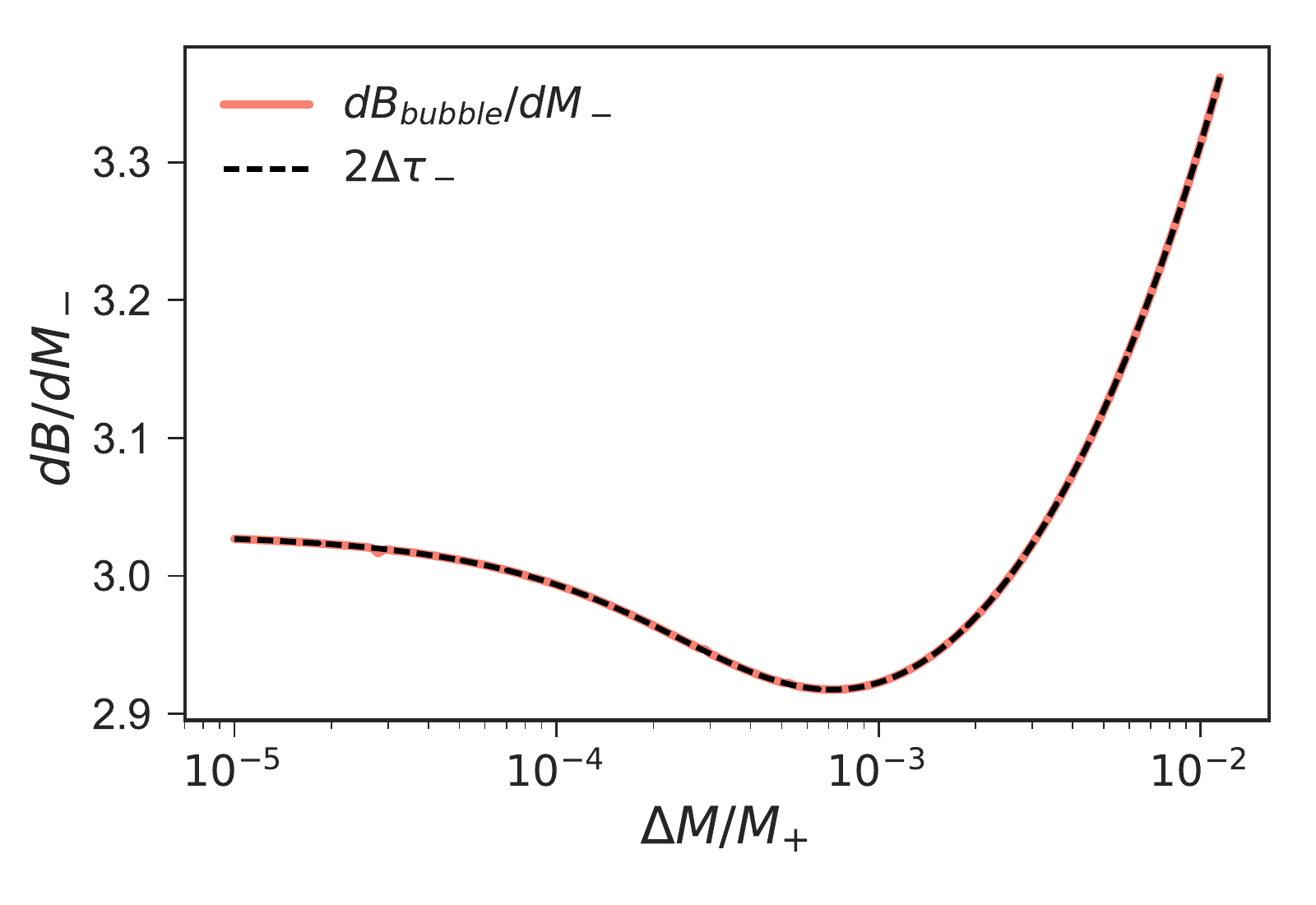} 
\caption{\small
 \textbf{Left}: $B_\text{bubble}$ (blue line) as a function of $\Delta M (\equiv M_+ - M_-)$, and
 \textbf{Right}: $\dd B_{\rm bubble} / \dd M_-$ (pink line) as a function of $\Delta M$,
 for a thin-wall bubble in the Schwarzschild-de Sitter spacetime. 
 In the left panel, we also plot $B$ (black thick line) and $B_{\rm boundary}$ (yellow dashed line); 
 in the right panel,  $2 \Delta \tau_-$ (black dashed) is shown.
 We take $R_{\text{BH},+} = 0.1 r_{\ast, 0}$, $\Lambda_+ = 0$, and $\Lambda_- =-0.03 / r_{\ast,0}^2$. 
 In this case, one can show that $l = 10 r_{\ast,0}$ and $\sigma l / M_\text{Pl}^2 = r_{\ast,0} / l = 0.1 \ll 1$.
 We take $r_{\ast,0} = 1 / M_\text{Pl}$ to plot the results, but 
 the $r_{\ast,0}$ dependence can be trivially factorized such that 
 $B, B_{\rm CdL} \propto r_{\ast,0}^2$, $\Delta M, M_+ \propto r_{\ast,0}$, 
 and $\dd B / \dd M_- \propto r_{\ast,0}$. 
 Note that $\dd B_{\rm bubble} / \dd M_- = - \dd B_{\rm bubble} / \dd \Delta M$. 
}
  \label{fig:B_E_BH}
\end{figure}

\begin{figure}[t]
\centering 
\includegraphics[width=.45\textwidth]{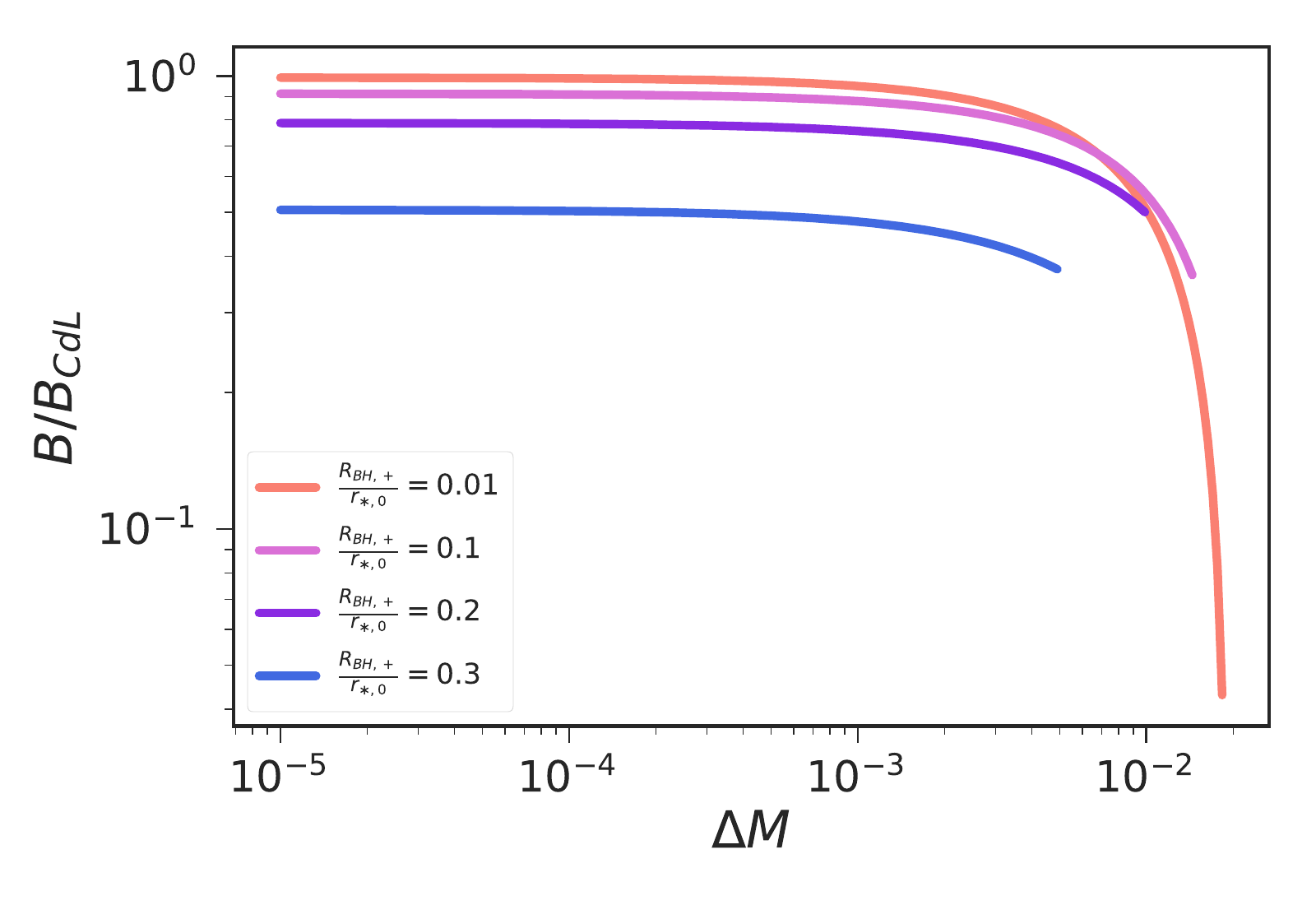} 
\qquad 
\includegraphics[width=.45\textwidth]{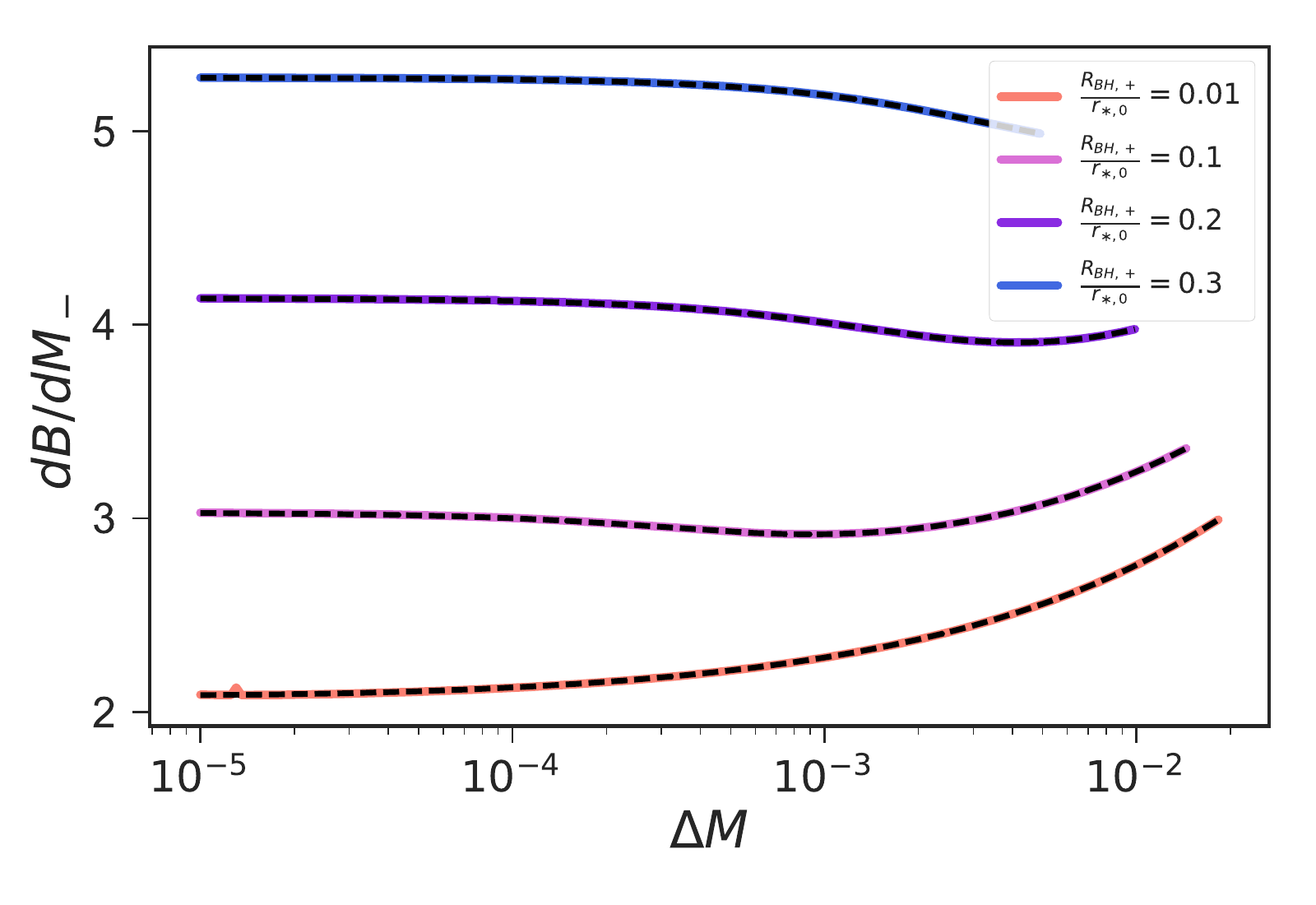} 
\caption{\small
 \textbf{Left}: $B$ as a function of $\Delta M (\equiv M_+ - M_-)$, and
 \textbf{Right}: $\dd B_{\rm bubble} / \dd M_-$ as a function of $\Delta M$,
 for a thin-wall bubble in the Schwarzschild-de Sitter spacetime. 
 We take the parameters to be the same as in Fig.~\ref{fig:B_E_BH} except that we take 
 $R_{\text{BH},+} /r_{\ast, 0}= 0.01$ (red line), $0.1$ (pink line), $0.2$ (violet line), 
 and $0.3$ (blue line). 
}
  \label{fig:B_E_varying}
\end{figure}

Let us take the variation of the bounce action
$B = B_\text{boundary} + B_\text{bubble}$ with respect to $M_-$
so as to approximate the integral of Eq.~\eqref{gamma with BH}.
The boundary term gives 
\begin{align}
 \frac{\dd }{\dd M_-} B_{\rm boundary} = 
 - \frac{4 \pi R_{\rm BH, -}}{1 - \Lambda_- R_{\rm BH,-}^2 } 
 = - T_{\rm BH, -}^{-1}. 
\label{B_boundary}
\end{align}
By numerically solving the equation of motion and calculating the transition rate
(See Figs.~\ref{fig:B_E_BH} and \ref{fig:B_E_varying}),
we also find that 
the variation of the other terms satisfies 
\begin{align}
 \frac{\dd }{\dd M_-} B_{\rm bubble} = 
 \int \dd \tau_- = 2 \Delta \tau_-. 
\end{align}
Combining these results, we obtain 
\begin{align}
 \frac{\dd }{\dd M_-} B = 
 2 \Delta \tau_- - T_{\rm BH, -}^{-1}. 
 \label{1/T formula for BH}
\end{align}

The above result \eq{1/T formula for BH} is similar to the one obtained in the previous sections [see \eq{T delta tau}]. 
In particular, 
if we require $\dd B (M_-) / \dd M_- = 0$, 
we find the relation $2 \Delta \tau_- = T_{\rm BH, -}^{-1}$, 
which may imply that the transition is due to the thermal effect with Hawking temperature.
However, the transition rate is not minimized at the saddle point 
unless $\dd^2 B(M_-) / \dd M_-^2 > 0$ as we discussed in the previous section. 
In fact, we numerically check that this condition is not always satisfied. 
We show an example in Fig.~\ref{fig:B_E_BH}, 
where we assume 
$R_{\text{BH},+} = 0.1 r_{\ast,0}$, $\Lambda_+ = 0$, and $\Lambda_- =-0.03 / r_{\ast,0}^2$.\footnote{
	Note that $r_{\ast,0} \equiv 3 \sigma / \Delta \epsilon$ ($\epsilon = \Delta \Lambda$).
}
These parameters lead to $l = 10 r_{\ast, 0}$ and $\sigma l / M_\text{Pl}^2 = r_{\ast,0}/l = 0.1$.
Note that the $r_{\ast,0}$ dependence can be trivially factorized in our results such that 
$B, B_{\rm CdL} \propto r_{\ast,0}^2$, $\Delta M, M_+ \propto r_{\ast,0}$, and $\dd B / \dd M_- \propto r_{\ast,0}$ though we take $r_{\ast,0} = 1/ M_\text{Pl}$ as a reference value to plot them. 
In the left panel, 
$B (\Delta M)$ monotonically decreases as $\Delta M$ increases 
and it is minimized at the maximal value of $\Delta M$, 
which corresponds to the static solution. 
Note again that, if the seed BH mass, $M_+$, is sufficiently heavy,
there exists a maximal value of $\Delta M$, above which we do not have solutions to Eq.~\eqref{EOM with BH}
and at which the solution becomes static.
In the right panel,  $\dd B_{\rm bubble}(\Delta M) / \dd M_-$ increases 
as $\Delta M$ increases for moderately large $\Delta M$, 
which implies that there is no saddle point at least in that range of $\Delta M$.

We can see that the result coincides with Fig.~\ref{fig:B_E} 
when we take a limit of $r_{\ast,0} \gg R_{\rm BH, +}$ (and $\Delta M \ll M_+$ as we discuss in the next subsection), where the effect of curved spacetime on the bubble configuration is negligible. 
Figure~\ref{fig:B_E_varying} shows the result of $B$ and its derivative with respect to $M_-$ 
for the cases of $R_{\text{BH},+} /r_{\ast, 0}= 0.01$ (red line), $0.1$ (pink line), $0.2$ (violet line), and $0.3$ (blue line). 
We can see that the red line in the figure, where we take $R_{\text{BH},+} /r_{\ast, 0} = 0.01 \ll 1$, 
coincides with that in Fig.~\ref{fig:B_E}. 
Note however that the result coincides with the one in the flat spacetime 
only when we identify $T_\ast$ in a flat spacetime 
as the Hawking temperature $T_{\rm BH, -}$ of the BH according to the correspondence 
between Eqs.~(\ref{T delta tau}) and (\ref{1/T formula for BH}). 
As we see below, this temperature is much larger than $T_{\rm *, th}$ defined by \eq{T_th}, 
so that the transition rate is dominated by the static solution. 
This is also true for larger values of $R_{\text{BH},+} /r_{\ast, 0}$, 
which then implies that the static solution dominates the transition rate 
even if $\dd^2 B_{\rm bubble} / \dd \Delta M_+^2 < 0$ for moderately small $\Delta M$ 
and large $R_{\text{BH},+} /r_{\ast, 0}$. 
These results show that the effect of the boundary term is relevant and is crucial 
even if the curved spacetime does not affect the bubble configuration.

\begin{figure}[t]
\centering 
\includegraphics[width=.45\textwidth]{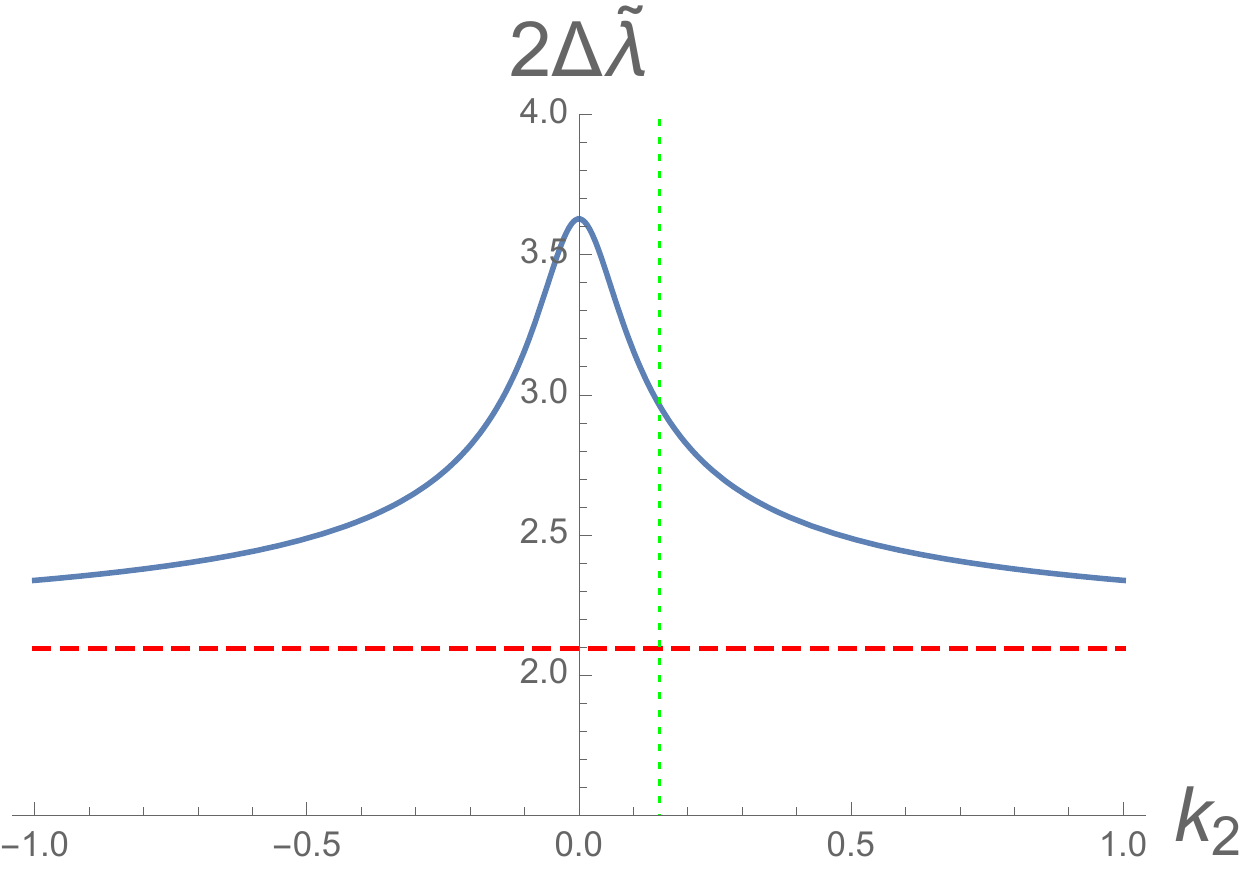} 
\caption{\small
 Relation between $2 \Delta \tilde{\lambda}$ and $k_2$ for static solutions. 
 The red dashed line represents the asymptotic value of $2 \Delta \tilde{\lambda}$ for $k_2 \to \pm \infty$. 
 The green dashed line represents an upper bound of $k_2$ for the case of $\Lambda_+ = 0$. 
}
  \label{fig:lambda-k2}
\end{figure}

Since the transition rate is smaller for smaller $M_-$, 
it is minimized at the critical point for a large initial BH mass. 
In fact, this property has been pointed out in Ref.~\cite{Gregory:2013hja} 
and so the static solution is studied in detail in the literature. 
To clarify the situation, 
we shall take the limit of $M_- = M_{-, \text{min}} + \varepsilon$ with $\varepsilon \to +0$,
and see what happens. 
$r_{*,1}$ and $r_{*,2}$ coincide with each other at $r_* = r_{*, \text{sp}}$, 
where 
\begin{align}
 &\tilde{r}_{*, \text{sp}} \equiv \sqrt{\frac{1}{9} \lmk 1 + A^{-1/3} + A^{1/3} \rmk } 
 \ \ \ge \frac{1}{3},
 \\
 &A \equiv 2^{-1/3} \lmk 2 + (27 k_2)^2 + 27 \abs{k_2} \sqrt{4 + (27 k_2)^2} \rmk, 
\end{align}
and the amplitude of the solution decreases to zero for $\varepsilon \to +0$. 
However, the second derivative $U''(\tilde{r})$ is nonzero in that limit. 
Since the potential can be approximated to be a quadratic potential for a sufficiently small $\varepsilon$, 
the periodicity is determined solely by the second derivative. 
It is given by 
\begin{align}
 &- \frac{\del^2 U}{\del \tilde{r}^2} 
 = 
 9 \lkk 1 - A^{1/3} + \frac{1}{ (27 k_2)^2} A^{1/3} 
 \lmk - B + \lmk B - (27 k_2)^2 \rmk A^{1/3} /2 \rmk \rkk, 
 \\
 &B \equiv  27 \abs{k_2} \sqrt{4 + (27 k_2)^2}. 
\end{align}
Note that $- (1/2)\del^2 U / \del \tilde{r}^2   = 3$ ($9$) for $k_2 = 0$ ($\pm \infty$). 
This is plotted in Fig.~\ref{fig:lambda-k2}, 
where the red dashed line represents the asymptotic value for $k_2 \to \pm \infty$. 
If $\Lambda_+ = 0$, 
there is an upper bound on $k_2$ indicated by the dotted line. 
We find that 
the periodicity of $\tilde{\lambda}$ is in the range of $(2 \pi / 3, 2 \pi / \sqrt{3})$ and is of order unity. 
Thus, the periodicity of $\tau_\pm$ for $M_- = M_{-,\text{min}} + \varepsilon$ with $1\gg \epsilon > 0$ 
is related to that of $\lambda$ as follows: 
\begin{align}
\lim_{\varepsilon\to +0}
 2 \Delta \tau_\pm (M_{-,\text{min}} + \varepsilon ) = \int \gamma_\pm \dd \lambda 
 = \frac{1}{\sqrt{f_\pm (r_{*,\text{sp}})}} \frac{\gamma_{\rm GMW}}{\alpha} 2 \Delta \tilde{\lambda}. 
\end{align}
Noting that $\gamma / \alpha \simeq r_{\ast, 0}$ for $\sigma^2 l^2 \ll 1$ 
and $\Lambda_+ = 0$, 
we find that the resulting periodicity is of the same order as the radius of the Coleman solution. 
Therefore, the periodicity of the bubble is much larger than the radius of the BH,
indicating that $\dd B / \dd M_- = 2 \Delta \tau_- - T_{\text{BH},-}^{-1}
\simeq 2 \Delta \tau_- > 0$ for $M_- = M_{-,\text{min}} + \varepsilon$ with $1 \gg \varepsilon > 0$.
This observation justifies that the transition rate tends to be minimized by the smallest value of $M_-$,
namely the static solution.
For a sufficiently small $M_\text{BH}$,
the minimum mass of the remnant BH becomes zero.\footnote{
	Such a seed mass is referred to as a critical mass in Refs.~\cite{Gregory:2013hja,Burda:2015yfa}.
}
In this special case, 
$\dd B / \dd M_- = \dd B_{\rm bubble} / \dd M_- = 2 \Delta \tau_-$, 
which indicates that the transition rate is minimized at $M_- = 0$ if possible.

Let us interpret the obtained result.
First, note that we can identify 
$\Delta M \equiv M_+ - M_-$ as the energy of the bubble because of the energy conservation. 
Then we can find 
some similarities between the above results and the ones in the previous section. 
First, 
the derivative of $B$ with respect to $\Delta M$ (or $- M_-$) 
gives the periodicity of the instanton solution and the inverse of the Hawking temperature of remnant BH $T_{\rm BH -}^{-1}$. 
This corresponds to \eq{T delta tau} once we identify the temperature as the Hawking temperature. 
Second, 
the second derivative tends to be negative 
(at least for a moderately large $\Delta M$ and/or $r_{\rm *,0} \gg R_{\rm BH, +}$).
Depending on the seed BH mass,
the transition rate is dominated by $M_- = 0$ or a static solution, 
which are the boundary of the allowed value of $M_-$. 
This is analogous to the thermal transition 
for a high enough temperature $T_\ast > T_{\ast, \text{th}}$
discussed in Sec.~\ref{sec:tunneling_thermal}.
(In fact, the Hawking temperature $T_{\rm BH, -}$ is much larger than 
the threshold temperature $T_{\rm *, th}$.) 
There, the periodicity of the bubble for $E < E_\text{sp}$ 
never coincides with $T_{\ast}^{-1}$.
This is not a problem because those solutions with $E < E_\text{sp}$
do not dominate the path integral in this case;
rather the static solution $E = E_\text{sp}$ is realized and
this bounce can be embedded into the Euclidean spacetime with the periodicity of $T_\ast^{-1}$.
We will further clarify these correspondences in the subsequent subsections.

\subsection{Bubble nucleation without backreaction to metric \label{sec3-2}}

In this section,
we neglect the change of the metric by the bubble in this subsection 
and regard the metric as a background~\cite{Basu:1991ig}.
Comparing the result computed in this way with the full gravitational one,
the meaning of the enhancement factor in Refs.~\cite{Gregory:2013hja, Burda:2015isa, Burda:2015yfa} is clarified.
In the thin-wall approximation, the worldsheet metric in the Schwarzschild--de Sitter spacetime background is written as 
\begin{align}
 \dd s^2_3 = - \lkk f(r_*) - \frac{1}{f(r_*)} \lmk \frac{\dd r_*}{\dd t} \rmk^2 \rkk \dd t^2 
 + r_*^2 \dd \Omega^2.
 \label{worldsheet metric2}
\end{align}
The Euclidean effective action for the domain wall is given by the 
worldsheet area in addition to the difference of potential energy inside and outside bubble, 
and thus we obtain 
\begin{align}
 S_E = \int \dd \tau \lkk 4 \pi r_*^2 \sigma \gamma^{-1} 
 - \frac{4 \pi}{3} (r_*^3 - R_{\rm BH}^3) \epsilon \rkk, 
 \label{S_E in SdS}
\end{align}
where $\epsilon = \Lambda_+ - \Lambda_-$ ($\equiv \Delta \Lambda$) and 
$R_{\rm BH} = M_{\rm BH} / 4 \pi$ is the BH horizon radius. 
Recall that the Euclidean gamma factor $\gamma$ is given by
\begin{align}
 \gamma = \frac{1}{\sqrt{f(r_*) + \frac{1}{f(r_*)} \lmk \frac{\dd r_*}{\dd \tau} \rmk^2}}. 
\end{align}
The conserved energy is 
\begin{align}
 E = \frac{\del L }{\del \dot{r}_*} \dot{r}_* - L, 
\end{align}
where the Lagrangian $L$ can be read from the above action. 
It can be expressed as
\begin{align}
 f(r_*) 4 \pi r_*^2 \sigma \gamma
 - \frac{4 \pi}{3} (r_*^3 - R_{\rm BH}^3) \epsilon  = E_*, 
 \label{EoM in SdS}
\end{align} 
where $E_*$ is the total energy in the initial state. 
We should take $E_* = 0$ 
for the vacuum transition without the thermal effect, 
in which case the bubble nucleates from $r_* = R_{\rm BH}$, \textit{i.e.}, from the surface of the BH, 
to a certain radius. 
The conservation law can be rewritten as 
\begin{align}
 \frac{1}{f^2(r_*)} \lmk \frac{\dd r_*}{\dd \tau} \rmk^2 = 
 f(r_*) \lkk \frac{\epsilon}{3 \sigma} r_* + \frac{E_*}{4 \pi r_*^2 \sigma} \rkk^{-2} - 1, 
 \label{Energy conservation law in SdS}
\end{align}
where we redefine $E_*  - 4 \pi R_{\rm BH}^3 \epsilon / 3$ as $E_*$. 

It is convenient to introduce the proper time $\lambda$ of the bubble trajectory: 
\begin{align}
 \dd \lambda = \gamma^{-1} \dd \tau, 
\end{align}
which gives 
\begin{align}
 f (r_*) \lmk \frac{\dd \tau}{\dd \lambda} \rmk^2 = 1 - \frac{1}{f(r_*)} \lmk \frac{\dd r_*}{\dd \lambda} \rmk^2. 
\end{align}
Then \eq{Energy conservation law in SdS} can be rewritten as 
\begin{align}
 \lmk \frac{\dd r_*}{\dd \lambda} \rmk^2 &= 
 f(r_*) - \lmk \frac{\epsilon}{3 \sigma} r_* + \frac{E_*}{4 \pi \sigma r_*^2} \rmk^2, 
 \\
 \frac{\dd \tau}{\dd \lambda} &=
 \frac{1}{f(r_*)} \lmk \frac{\epsilon}{3 \sigma} r_* + \frac{E_*}{4 \pi \sigma r_*^2} \rmk. 
 \label{eq:master_wo_BH}
\end{align}
Recalling that 
\begin{align}
 \lkk 1 - \frac{k_1}{\tilde{r}_\ast} - \lmk \tilde{r}_\ast + \frac{k_2}{\tilde{r}_\ast^2} \rmk^2 \rkk 
 \lkk \frac{\tilde{r}_\ast}{\alpha} + \frac{k_2}{\tilde{r}_\ast^2} \rkk^{-2} 
 = 
 f_-(r_\ast) \lkk \frac{k_2}{\tilde{r}_\ast^2} + \frac{\tilde{r}_\ast}{\alpha} \rkk^{-2}   - 1, 
\end{align}
we find that this equation is consistent with the result in the previous subsection 
in the limit of $f_- (r) \approx f(r)$, $\gamma_{\rm GMW} \approx 3 \sigma / \epsilon = r_{\ast,0}$, 
and $\Delta M  =  E_*$. 
Here, extra terms in $\gamma_{\rm GMW}$ other than $3 \sigma / \epsilon$ in Refs.~\cite{Gregory:2013hja,
Burda:2015isa, 
Burda:2015yfa, Burda:2016mou} come from 
the backreaction to the bubble, as we see below.
Thus our result is consistent with their result in the limit of the negligible backreaction.

Here we would like to explain that the exact form of $\gamma_{\rm GMW}$
can be understood as the backreaction to the bubble.
In the next-to-leading order approximation, 
we have to take into account the self-gravitational energy of the bubble 
in the Newtonian limit. 
It may be calculated by 
\begin{align}
 E_{\rm self} 
 &=
 \frac{1}{2} 
 \int r_*^2 \dd \Omega^{'2} 
 \int r_*^2 \dd \Omega^2 \frac{G \sigma \sigma}{\abs{{\bf r - r'}}} 
 \\ 
 &=
 \pi r_*^3 \sigma^2, 
\end{align}
where $G = 1/(8 \pi)$ is the Newton constant. 
When we include this contribution to the bubble action, 
it may be absorbed by the redefinition of $\epsilon$, 
\begin{align}
 \epsilon \to \epsilon \lmk 1 - \frac{3 \sigma^2}{4 \epsilon} \rmk.
\end{align}
This is because the $\epsilon$ dependence comes only through $\gamma_{\rm GMW} \simeq 3 \sigma / \epsilon$ in 
the previous leading order results.
Then the next-to-leading order contribution 
is
\begin{align}
 \gamma_{\rm GMW} \simeq 
 \frac{3\sigma}{\epsilon} \lmk 1 - \frac{3\sigma^2}{4 \epsilon} \rmk, 
\end{align}
which is consistent with the expansion of the exact $\gamma_{\rm GMW}$ under $\sigma^2 / \epsilon \ll 1$.

Now we are ready to discuss the bubble nucleation with a finite energy under this metric,
and see its relation to the result in the previous section.
First we need to derive the bubble nucleation rate with an initial energy of $E_*$, 
where the bubble solution obeys \eq{EoM in SdS}. 
It is given by 
\begin{align}
 B (E_*) 
 &= S_{\rm bounce} (E_\ast) - S_{\text{E},0} (E_\ast) 
 \\
 &= \int \dd r_* \sqrt{ \frac{1}{f(r_*)} \lmk 4 \pi r_*^2 \sigma \rmk^2 -  \frac{1}{f^2(r_*)} \lmk \frac{4}{3} \pi r_*^3 \epsilon + E_* \rmk^2} 
 \\
 &= \int \dd \tau 4 \pi r_*^2 \sigma \gamma \frac{1}{f(r_*)} \lmk \frac{\dd r_*}{\dd \tau} \rmk^2,  
 \label{B_b in SdS}
\end{align}
with $S_{\text{E},0} (E_\ast) = E_* \Delta \tau$. 
To get the vacuum decay rate,
we have to multiply the probability of producing a bubble with $r_{\ast,1}$.
Let us consider the situation
where bubbles with $E_{\ast}$ are produced with
a rate of $e^{- P(E_\ast)}$.
Then
the bubble nucleation rate for $E_\ast < E_\text{sp}$ is given by 
\begin{align}
 \Gamma_q \left( E_*; [P] \right) \sim e^{- P (E_\ast)} e^{-B (E_*)}. 
 \label{gamma_q M-}
\end{align}

\begin{figure}[t]
\centering 
\includegraphics[width=.45\textwidth]{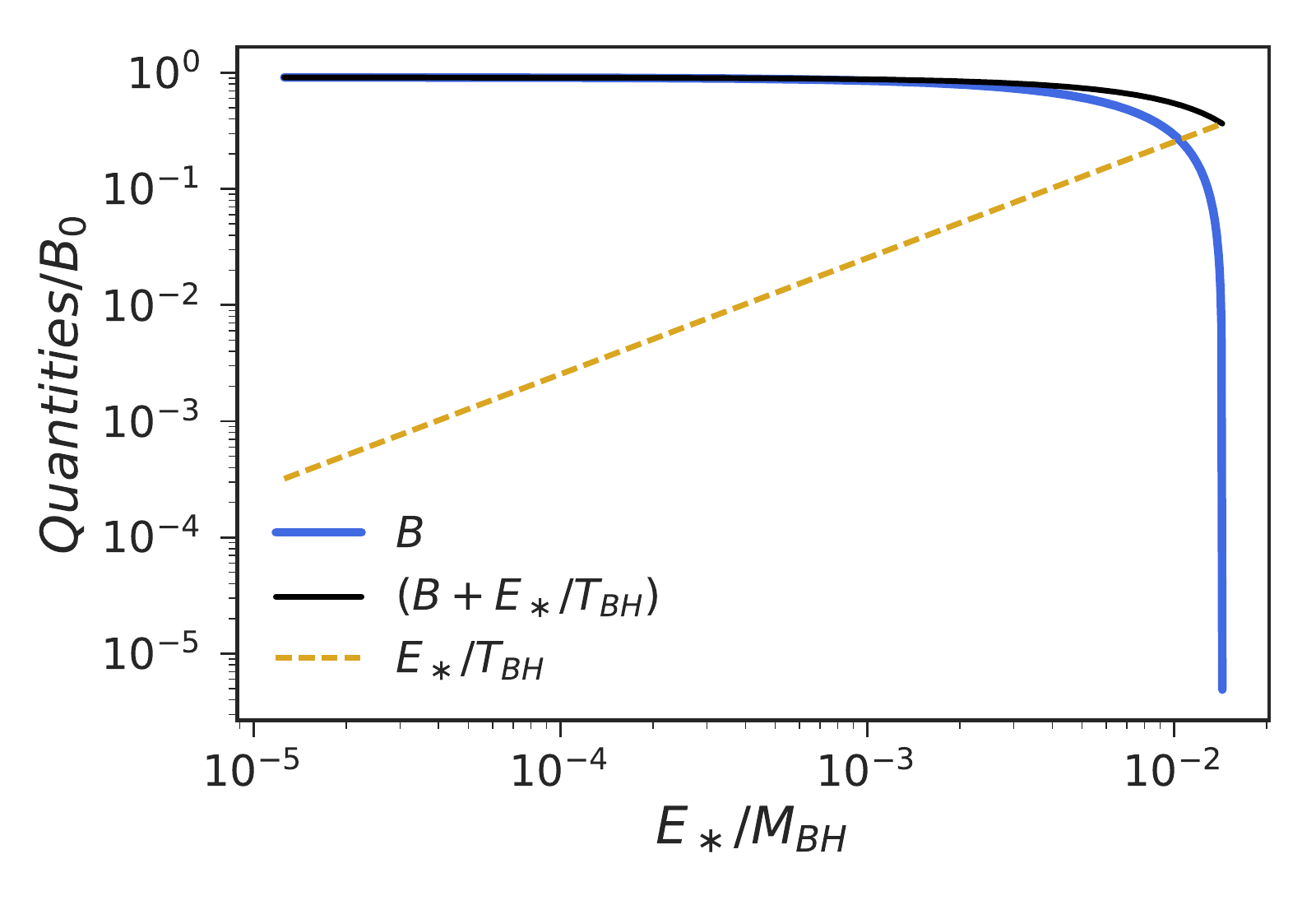} 
\qquad 
\includegraphics[width=.45\textwidth]{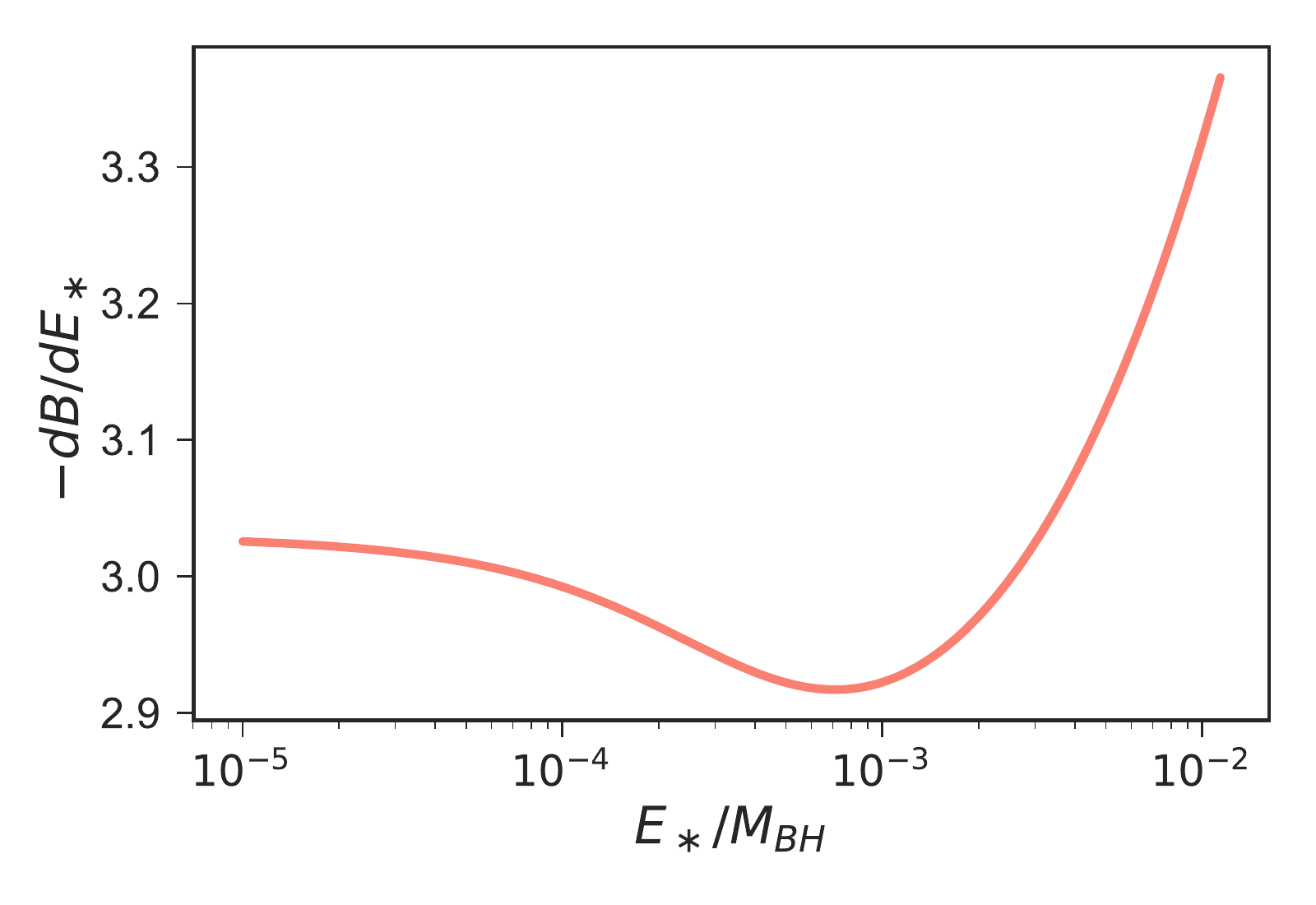} 
\caption{\small
 Plot of $B(E_*)$ (left panel) and $-\dd B(E_*) / \dd E_*$ (right panel) as a function of $E_*$ 
 for a thin-wall bubble in the Schwarzschild spacetime
 without the backreaction  of the bubble to the metric.
 In the left panel, we also plot $B + E_\ast / T_\text{BH}$ (black thick line) and $E_\ast / T_\text{BH}$ (yellow dashed line).
 We take $R_{\text{BH},+} = 0.1 r_{\ast,0}$ 
 and $l = 10 r_{\ast,0}$. 
 We also take $r_{\ast,0} = 1 / M_\text{Pl}$ 
 though the $r_{\ast, 0}$ dependence can be trivially factorized 
 as $\dd B / \dd E_* \propto r_{\ast, 0}$. 
 One can see that the result almost coincides with Fig.~\ref{fig:B_E_BH}.
}
  \label{fig:wo_backreaction}
\end{figure}

\begin{figure}[t]
\centering 
\includegraphics[width=.45\textwidth]{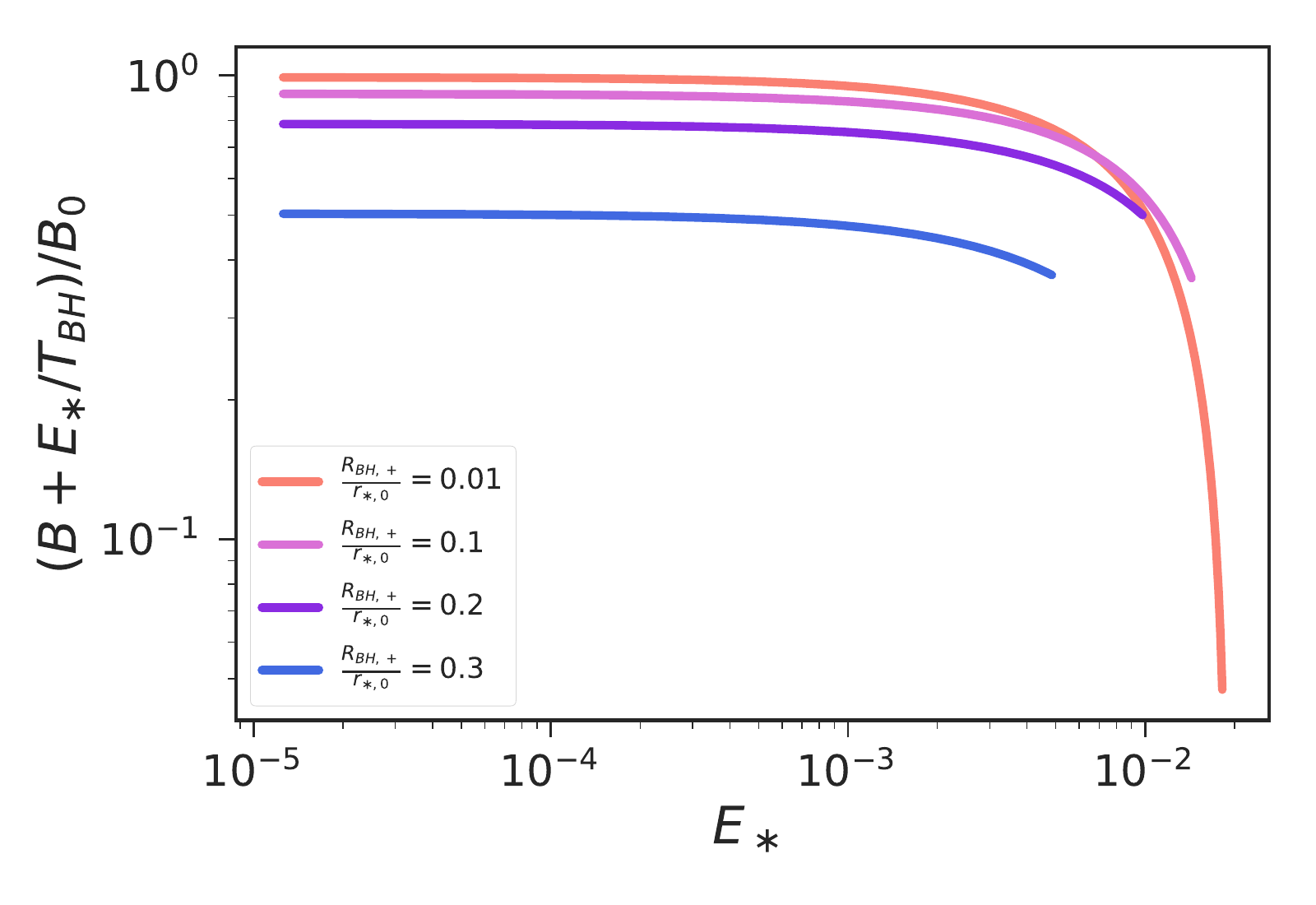} 
\qquad 
\includegraphics[width=.45\textwidth]{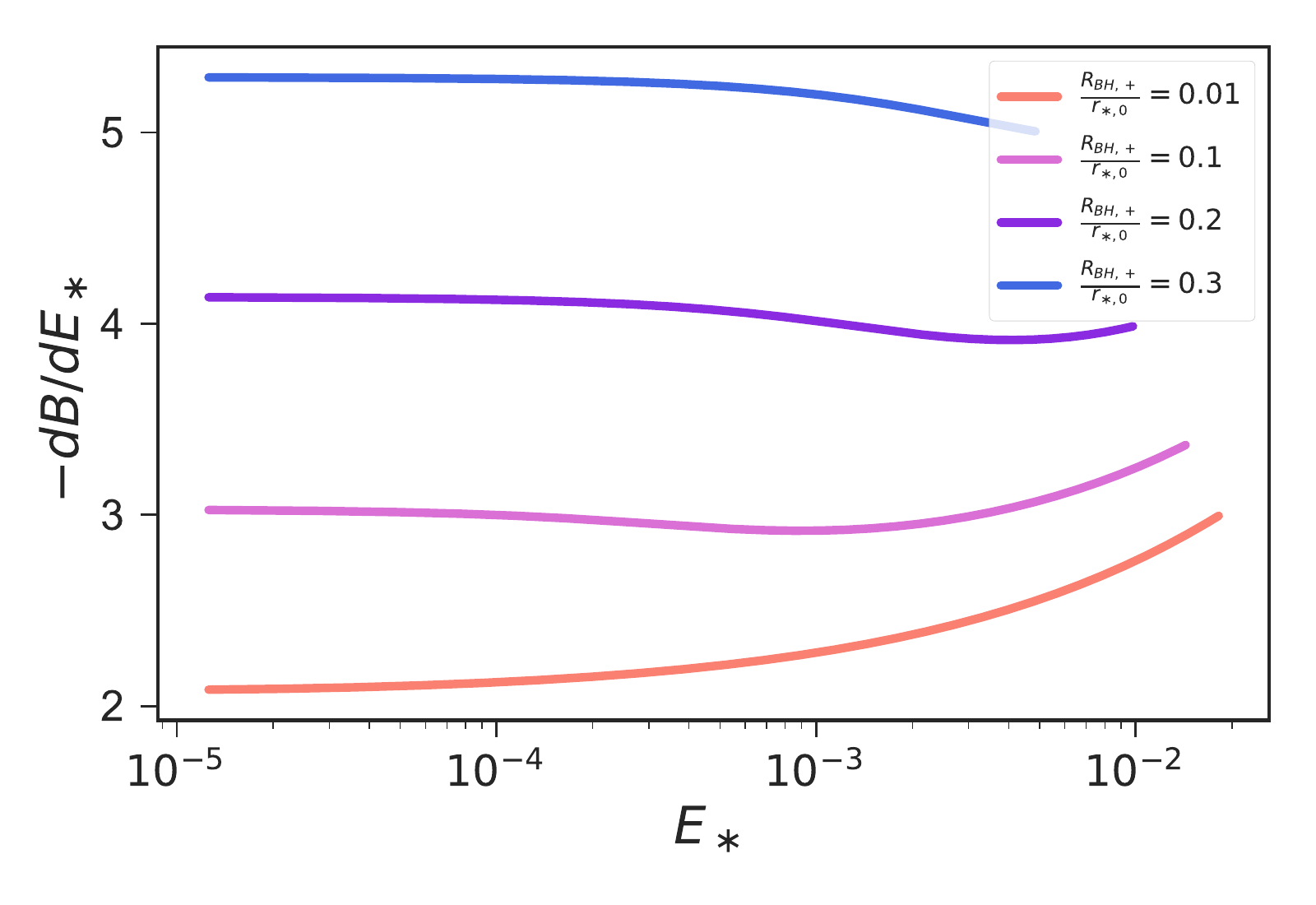} 
\caption{\small
 Plot same as Fig.~\ref{fig:wo_backreaction}
 but with $R_{\text{BH},+} /r_{\ast, 0}= 0.01$ (red line), $0.1$ (pink line), $0.2$ (violet line), and $0.3$ (blue line). 
 One can see that the result almost coincides with Fig.~\ref{fig:B_E_varying}.
}
  \label{fig:wo_backreaction_varying}
\end{figure}

Up to here, the probability function $P(E)$ is generic.
In the following, we would like to specify its form
so that it reproduces the previous result in a certain limit.
Then, we discuss its consequence.
First, we show that
$B (E_\ast)$ coincides with $B_\text{bubble}$ [Eq.~\eqref{eq:bubble_BH}]
in the limit of $f_- \approx f$, $\gamma_\text{GMW} \approx 3 \sigma / \epsilon = r_{\ast,0}$
and $\Delta M = E_\ast$.
This is obvious because we have already seen that the master Eq.~\eqref{eq:master_wo_BH}
coincides with that in the previous section [Eqs.~\eqref{EOM with BH} and \eqref{eq:EOM_potential_BH}].
For clarity, we also plotted $B(E_\ast)$ and $-B'(E_\ast)$ in Fig.~\ref{fig:wo_backreaction} 
and Fig.~\ref{fig:wo_backreaction_varying} 
for parameters which are close to the case shown in Fig.~\ref{fig:B_E_BH}
and Fig.~\ref{fig:B_E_varying}, respectively.
One can see that $B(E_\ast)$ and $-B'(E_\ast)$ coincide with those in Fig.~\ref{fig:B_E_BH} 
and Fig.~\ref{fig:B_E_varying},
since, for the parameters in these figures, the backreaction is quite small
because of $\sigma l = r_{\ast,0}/l = 0.1 \ll 1$ and $M_+ \gg E_\text{sp} \geq \Delta M$.
Next, let us see how $B_\text{boundary}$ behaves in this limit.
Since we have $M_\text{BH} \approx M_+ \approx M_- \gg \Delta M$ 
and $1 \gg r_{\ast,0}/l \gtrsim M_- / 4 \pi l$,
the boundary term can be expressed as
\begin{align}
	B_\text{boundary} &= \frac{1}{2} \left[ M_+^2
	- M_-^2 \left(1 +  \frac{(M_-/4\pi)^2}{l^2} + \cdots \right)^2 \right]
	\approx \frac{\Delta M}{1/M_\text{BH}} = \frac{E_\ast}{T_\text{BH}}.
\end{align}
In the last equality, we use $T_\text{BH} = 1 / M_\text{BH}$
and identify $E_\ast = \Delta M$.
Now it is clear that the result in the previous section is reproduced for
the probability function of
\begin{align}
	P (E_\ast) = \frac{E_\ast}{T_\text{BH}},
	\label{P(E)}
\end{align}
which is nothing but the canonical ensemble with a temperature $T_\ast$
that coincides with the Hawking temperature of the background BH, $T_\ast = T_\text{BH}$.

Before discussing the physical meaning of $P(E_\ast)$, let us evaluate the vacuum decay rate.
It is given by
\begin{align}
	\Gamma = \int \dd E_\ast  \Gamma_q \left( E_*; [P] \right) \sim \int \dd E_\ast e^{- P (E_\ast)} e^{- B (E_\ast)}.
\end{align}
Again, let us first try to find the saddle point. A first derivative of the exponent yields
\begin{align}
	\frac{\dd P}{\dd E_\ast} &= T_\text{BH}^{-1},\\
	- \frac{\dd B}{\dd E_\ast} &= \int \dd \tau \frac{(4/3) \pi r_\ast^3 \epsilon + E_\ast}{f(r_\ast) 4 \pi r_\ast^2 \sigma \gamma} 
	= \int \dd \tau = 2 \Delta \tau.
\end{align}
If the second derivative of the bounce with respect to $E_\ast$ is always positive,
one may approximate the integral by the saddle point satisfying $2 \Delta \tau = T_\text{BH}^{-1}$.
However, the second derivative can be negative in some cases (see Fig.~\ref{fig:wo_backreaction} and Fig.~\ref{fig:wo_backreaction_varying}).
Actually, the sign is always negative in the quantum field theory in the flat spacetime 
as we have discussed in Sec.~\ref{sec:tunneling_thermal}. 
If $B''(E)$ is negative for $0 \leq E_\ast \leq E_\text{sp}$, the transition rate may not be dominated by the saddle point 
but by either boundary of $E_*$. 
One of the boundaries of $E_*$ is, of course, $E_* = 0$, 
which is just the quantum tunneling process from a false vacuum in the presence of a BH with $M_\text{BH}$~\cite{Hiscock:1987hn}. 
The other boundary $E_{\rm sp}$ comes from the condition that $B(E_{\rm sp}) = 0$, 
which corresponds to the sphaleron process at high temperature 
and is determined by the static solution. 
For the sample parameters shown in Fig.~\ref{fig:wo_backreaction} and Fig.~\ref{fig:wo_backreaction_varying},
one can see that $E_\ast = E_\text{sp}$ dominates for $T_{*} = T_{\rm BH, -}$.
Here we assume $E_* \ll M_+$ so that we can neglect the change of metric 
due to the bubble nucleation. Hence we require $E_{\rm sp} \ll M_+$. 
Note again that the dominant process is always consistent with the periodicity indicated by
the BH Hawking temperature $T_\text{BH}$,
although a family of bounce solutions as a function of $E_\ast$ is not.

Now we would like to discuss the consequence of $P(E_\ast)$.
Comparing the bubble nucleation rate computed in this section with the full gravitational one,
we have seen that the enhancement factor can be regarded as a probability of generating
bubbles with a finite energy $E$.
However, there are many other degrees of freedom in quantum field theory,
and hence it is hard to imagine that a BH only excites bubbles.
Therefore, all the states with an energy $E$ other than bubbles should also be generated by the same mechanism.
This observation makes it clear that one needs to take into account finite-density corrections to the bubble nucleation rate induced by the presence of plasma.
Up to here, the conclusion is independent of whether or not the plasma fills the whole Universe.
However, the size of corrections does depend.
If $P(E_\ast)$ originates from Hawking radiation emitted from the BH horizon,
the flux decreases as we move away from the BH.
If it originates from the thermal plasma of the Hawking temperature filling the Universe,
the thermal plasma is present even far away from the BH.
We expect that the finite-density corrections in the former case are milder than those in the latter.
See Sec.~\ref{conclusions} for further discussion.

\subsection{Bubble nucleation via sphaleron}\label{sec:sphaleron}

At a sufficiently high temperature, 
the classical transition rate $\Gamma_c$ dominates over the transition rate~\cite{Hawking:1981fz}, 
which corresponds to a static solution to \eq{EoM in SdS}. 
Then, the transition rate is given by $e^{-E_{\rm sp} /T_*}$, 
where $E_{\rm sp}$ is the bubble energy for the static solution. 
This is nothing but the sphaleron transition. 
In this subsection, we study this sphaleron transition 
in more detail.

\subsubsection{de Sitter spacetime}
\label{sphaleron de Sitter}

First, let us focus on the Hawking-Moss transition in the de Sitter spacetime, 
where the metric in the static patch can be written as 
\begin{align}
 \dd s^2 &= - f_{\rm dS} (r) \dd t^2 + \frac{\dd r^2}{f_{\rm dS}(r)} + r^2 \dd \Omega,
 \\
 f_{\rm dS}(r) &= 1 - \frac{\Lambda r^2}{3}. 
 \label{dS metric}
\end{align}
This coordinate has an apparent horizon with a radius of $H^{-1} = \sqrt{3 / \Lambda}$. 

Suppose that the transition occurs in the whole region inside the horizon, 
in which case 
we should take into account the boundary term of the gravitational action. 
Since the scalar field is static inside the bubble 
and the whole region is contained inside the bubble, 
the solution is static and 
the transition rate is dominated by the boundary term: 
\begin{align}
 B = 8 \pi^2 \lmk \frac{3}{V (\phi)} - \frac{3}{V_{\rm FV}} \rmk, 
 \label{sphaleron in de Sitter}
\end{align}
where $V(\phi)$ is the local maximal of the scalar potential. 
In gravity theory, the static solution results in $H=0$ (\textit{i.e.}, $E = 0$) 
by the Hamiltonian constraint, 
which is the consequence of the least action principle for Laps function and 
some other components in the metric~\cite{Hawking:1995fd}. 
Therefore, 
the tunneling rate is determined only by the change of the boundary term $\Delta S_T$ for the static solution 
as \eq{sphaleron in de Sitter}~\cite{Oshita:2016oqn},
analogous to the thermodynamic transition 
with a conserved energy, like a \textit{microcanonical} picture.

A similar result can be derived when we neglect the effect of the change of metric 
and regard the metric as a background. 
This case is analogous to the thermodynamic transition
with unchanged temperature, like a \textit{canonical} picture,
as we will see.
First, note that we use the metric of \eq{dS metric}, 
which has an apparent boundary at $r_H = H^{-1}$. 
It can be rewritten as 
\begin{align}
 \dd s^2 = 
 - \dd T^2 + \dd S^2 + \dd X^2 + \dd Y^2 + \dd Z^2, 
\end{align}
with 
\begin{align}
 - T^2 + S^2 + X^2 + Y^2+ Z^2 = H^{-2}, 
 \\
 T = \sqrt{H^{-2} - r^2 } \sinh \lkk H t \rkk, 
 \\
 S = \sqrt{H^{-2} - r^2 } \cosh \lkk H t \rkk, 
 \\
 X = r \sin \theta \cos \phi, 
 \\
 Y = r \sin \theta \sin \phi, 
 \\
 Z = r \cos \theta. 
\end{align}
The metric has a singularity at $T = S = 0$ unless 
the time variable $t$ is periodic in the imaginary part with a period of $T^{-1}_{\rm dS} = 2\pi / H$. 
Therefore, for an equilibrium state in de Sitter space, 
the propagator of a scalar field has a periodicity in $i(t - t')$ with a period of $T^{-1}_{\rm dS} = 2\pi / H$, 
which implies that 
the scalar field is mimicked as in a thermal system with a temperature of $T_{\rm dS} = H / 2 \pi$.\footnote{
	Unless it has a conformal coupling $\xi = 1/6$.
}
Thus the scalar field has thermal fluctuation 
and in fact 
Hawking and Moss have pointed out that 
the thermal transition of the Universe inside the horizon 
can occur due to the thermal effect of Hawking radiation in de Sitter spacetime. 
When we consider a nucleation of a bubble with the size of Hubble volume, 
it is static and 
its energy is given by 
\begin{align}
 E_{\rm sp} = \frac{4 \pi}{3} H^{-3} \Delta V (\phi). 
 \label{sphaleron energy in de Sitter BG}
\end{align}
The static bubble is nucleated by the following sphaleron rate: 
\begin{align}
 \Gamma_{\rm sp} \propto e^{-E_{\rm sp}/T_{\rm dS}}. 
 \label{sphaleron in de Sitter BG}
\end{align}
This transition process is just the Hawking-Moss transition 
and the exponential factor is equal to the result [\eq{sphaleron in de Sitter}] 
in the limit of $\Delta V \ll V$ 
($\lkk V (\phi) - V_{\rm FV} \rkk / V^2 (\phi) \simeq 1/V(\phi) - 1/ V_{\rm FV}$). 
This equivalence also implies that 
the above calculation of \eq{sphaleron in de Sitter}, 
where we include the boundary term as proposed in Refs.~\cite{Gregory:2013hja,
Burda:2015isa, 
Burda:2015yfa}, 
corresponds to the transition due to the thermal effect 
with the Hawking temperature.

Next, let us consider the same theory 
with a metric which can describe the outer region of the boundary~\cite{Starobinsky:1994bd}. 
This result can be reproduced by calculating a distribution function of $\phi$ 
for a mode larger than the Hubble horizon scale. 
The metric is written by 
\begin{align}
 \dd s^2 = \dd t^2 - a_0^2 e^{2 H t} \dd x^2
 = ( H \eta)^{-2} \lmk \dd \eta^2 - \dd x^2 \rmk, 
 \label{nohorizon metric}
\end{align}
where $\eta$ is the conformal time defined by $a_0 e^{Ht} = (-H\eta)^{-1}$ 
for $- \infty < \eta < 0$. 
Let us first 
assume that the scalar field is massless, in which case 
the equation of motion is given by 
\begin{align}
 \ddot{\phi}_k + 3 H \dot{\phi}_k + \frac{k^2}{a^2(t)} \phi_k = 0, 
\end{align}
in the Fourier space. 
The solution is given by 
\begin{align}
 \phi_k 
 = \sqrt{\frac{\pi}{4}} H ( - \eta)^{3/2} H_{3/2}^{(1)} (-k \eta) 
 = \frac{H}{\sqrt{2k}} \lmk \eta - \frac{i}{k} \rmk e^{-ik\eta}, 
\end{align}
where $H^{(1)}_{3/2}$ is the Hankel function of the first class. 
Here, we implicitly assume that 
the vacuum state annihilated by $a_k$'s corresponds to the usual adiabatic vacuum in the limit of $\eta \to - \infty$. 
Such a vacuum is chosen in the context of inflation 
and corresponds to the one naturally defined in the metric \eq{dS metric}. 
This is because well inside the horizon the scalar field is just a free theory one, 
which is equivalent to the one without metric (\textit{i.e.}, Minkowski metric). 
When we focus on a mode with the wavelength of the order of the Hubble length in a de Sitter vacuum, 
the particle distribution function $\rho (\phi, t)$ is calculated as 
\begin{align}
 \rho (\phi, t) = 
 N^{-1} 
  \exp \lmk - \frac{8 \pi^2 }{3 H^4} \Delta V(\phi) \rmk, 
  \label{SY}
\end{align}
where $N$ is a normalization factor~\cite{Starobinsky:1994bd}. 
Therefore, 
the probability for a transition to $\phi_2$ is given by 
\begin{align}
 \Gamma \propto \exp [ - 8 \pi^2/(3 H^4) ( V(\phi_2) - V_{\rm FV}) ]. 
\end{align}
Since $T_{\rm dS} = H / (2 \pi)$, 
this result is consistent with the above ones.

Here, we summarize these methods and results. 
When we consider the metric with a boundary at $r = H^{-1}$ 
and regard it as a background, 
the Hawking-Moss transition occurs due to the thermal effect [see \eq{sphaleron energy in de Sitter BG}]. 
When we consider the same metric and take into account the Einstein equation 
as well as the equation of motion for the scalar field, 
the transition is a static solution 
and 
its rate is given by the surface term (or entropy) [see \eq{sphaleron in de Sitter}]. 
When we consider the metric without a boundary and use the Bunch-Davis (BD) vacuum, 
the transition can be calculated from the method used by Starobinsky and Yokoyama [see \eq{SY}]. 
These calculations are equivalent 
because the initial state is the same in each.

We may obtain some insights into the origin of the thermal plasma of the Hawking temperature 
in the above calculation 
by noting that there are ambiguities in the meaning of a vacuum state 
in curved spacetime~\cite{Unruh:1976db, Gibbons:1976ue, Gibbons:1977mu, Bunch:1978yq}. 
There is no apparent boundary in the metric \eq{nohorizon metric}, 
so it is natural to consider a nonsingular vacuum. 
It is actually the case for the BD vacuum 
$\left\vert 0_{\rm BD} \right\rangle$~\cite{Bunch:1978yq}. 
Then the transition rate is calculated from 
\begin{align}
 e^{- \Gamma t_0} 
 &=
 \frac{\abs{\la \phi_{\rm bubble}, \tau = \tau_0 \vert \phi_{\rm FV}, \tau = 0, {\rm BD} \ra}^2}
 {\abs{\la \phi_{\rm FV}, \tau = \tau_0, {\rm BD} \vert \phi_{\rm FV},\tau = 0, {\rm BD} \ra}^2}
 \\
 &=
 \int \mathcal D \phi e^{i S [\phi] - i S[\phi_{\rm FV}]}, 
\end{align}
as explained above. 
Then, after the coordinate transformation to the metric \eq{dS metric}, 
the imaginary time should be periodic so as to avoid the singularity at the apparent horizon. 
Since the periodicity of imaginary time effectively leads to a thermal effect on the scalar field, 
the BD vacuum state 
is seen as a thermally excited state by the observer using the latter metric. 
Therefore, for the latter observer, 
the transition rate from the same state 
is schematically written as 
\begin{align}
 e^{- \Gamma t_0} 
 &=
 \frac{\abs{\sum_{E, i} \la \phi_{\rm bubble}, \tau = \tau_0 \vert E, i \ra \la E, i \vert \phi_{\rm FV}, \tau = 0, {\rm BD} \ra}^2}
 {\abs{\la \phi_{\rm FV}, \tau = \tau_0, {\rm BD} \vert \phi_{\rm FV}, \tau = 0 , {\rm BD} \ra}^2 }, 
\end{align}
where we insert a complete set $\sum_{E, i} \left\vert E, i \ra \la E, i \right\vert = 1$. 
Since the BD vacuum is a thermally excited state with the temperature of $T_{\rm dS} = H / (2 \pi)$ 
from the observer with the metric with the apparent boundary, 
we use $\la E, i \vert \phi_{\rm FV}, \tau = 0, {\rm BD} \ra \sim e^{- E / T_{\rm dS}}$ 
and rewrite it as 
\begin{align}
 \Gamma 
 \sim \int_{0}^{V_{\rm top}} \dd E 
 e^{- E/ T_{\rm dS}} e^{-B(E)} 
 + 
 \int_{V_{\rm top}}^\infty \dd E 
 e^{-E / T_{\rm dS}}. 
\end{align}
Note that we also have to include the boundary term in $B(E)$ because 
there is the apparent singularity. 
Of course, 
these results represent the nucleation rate from the same state $\left\vert \phi_{\rm FV}, \tau = 0, {\rm BD} \ra$, 
so they should be equal to each other. 

This example implies that the effective thermal effect is implicitly included when we calculate transition rates in a Euclidean background metric. This should also be true in the Schwarzschild--de Sitter spacetime. In fact, it has been discussed that a no-boundary wave function of BH in the Euclidean geometry coincides with the Hartle-Hawking vacuum state~\cite{Hartle:1976tp, Barvinsky:1994jca}. This fact may indicate that the bubble nucleation rate is enhanced by the thermal effect of the Hawking radiation.

\subsubsection{Schwarzschild--de Sitter spacetime}
\label{SdS sphaleron}

Now, we move to the sphaleron transition in the Schwarzschild--de Sitter spacetime.

Let us first take the viewpoint of Sec.~\ref{sec3-1}.
There may be no solution to \eq{EOM with BH} for small values of $M_-$ 
when $M_+$ is larger than a certain threshold value~\cite{Gregory:2013hja,
Burda:2015isa, 
Burda:2015yfa, Burda:2016mou}. 
In particular, this is the case for $k_1 \gg 1$ and/or $k_2 \gg 1$. 
In this case, 
the transition rate is dominated by a static solution 
and 
the sum of the matter and the gravitational action vanishes 
due to the Hamiltonian constraint except for the boundary term. 
Thus the transition rate 
is just given by the boundary term in $\Delta S_G$ 
as $2 \pi \Delta A  = (M_+^2 - M_-^2) / 2$. 
Note that since the total energy is conserved in this calculation, 
the bubble energy is equal to the change of the BH mass: 
$\Delta E_\text{bubble} = \Delta M$. 
Then the bounce action can be interpreted as a 
change of the entropy associated with the Hawking temperature $\dd S = \dd U / T_{\rm BH, -}$ 
because 
\begin{align}
B_{\rm boundary} = \int \dd M_- \frac{\dd}{\dd M_-} B_{\rm boundary} 
= - \int \frac{\dd M_-}{T_{\rm BH,-} }
= \int \dd S, 
\end{align}
where we use \eq{B_boundary}. 
One can see that this viewpoint is analogous to 
the thermodynamic transition where the energy is conserved
during that process, like a microcanonical picture.

Then, let us discuss the viewpoint of Sec.~\ref{sec3-2}.
There may be no solution satisfying the equation of motion \eq{Energy conservation law in SdS} 
for the energy larger than a certain threshold value. 
This implies that the transition occurs classically 
with the probability function of \eq{P(E)} for such a large initial energy. 
Thus, we obtain 
\begin{align}
 \Gamma_{\rm BH} \sim 
 \int^{E_{\rm sp}}_0 
 \dd E_* e^{- P(E_\ast) - B(E_\ast)}
 + \int_{E_{\rm sp}} \dd E_* e^{-P(E_*)}, 
\end{align}
where the first term is just \eq{gamma_q M-} 
and the second term is the classical contribution. 
The sphaleron energy $E_{\rm sp}$ is given by 
the bubble energy for the static solution. 
It is clear that we have the classical transition process 
from this viewpoint. 
When we use \eq{P(E)} and consider a system with a sufficiently high temperature, 
the classical transition rate $\Gamma_c$ dominates over the transition rate in \eq{gamma_q and c}, 
which corresponds to a static solution to \eq{EoM in SdS}. 
In this case, 
the transition rate is given by $e^{-E_{\rm sp} /T_*}$, 
where $E_{\rm sp}$ is the bubble energy for the static solution. 
This viewpoint is analogous to the thermodynamic transition
where the temperature is unchanged instead of the energy,  
like a canonical picture.

Note that these two pictures give the same result in a certain limit,
as in the case of the equivalence between the microcanonical and canonical ensembles.
In the viewpoint of Sec.~\ref{sec3-1}, 
the energy is conserved $E_\text{sp} = \Delta M $.
Thus 
the transition rate of the sphaleron process in Sec.~\ref{sec3-2}  
is interpreted as to $e^{-\Delta M / T_*}$. 
This can be farther rewritten as $e^{- M_{\rm BH} \Delta M} \simeq e^{- (M_+^2 - M_-^2) / 2} = 
e^{- 2 \pi \Delta A}$, 
where we use $M_+ \simeq M_-$,
which should be satisfied to match the result in Sec.~\ref{sec3-2}
regarding the metric as a background.
Thus, both results coincide if
the backreaction of the bubble energy to the metric can be neglected. 
This result also implies that 
the transition occurs via the thermal fluctuation of the Hawking temperature.

\section{Conclusions and Discussion
\label{conclusions}}

In this paper, we reconsidered the bubble nucleation around a BH by using an effective theory of a thin-wall bubble in the Schwarzschild-de Sitter spacetime. 
We calculated the bubble nucleation rate in the Schwarzschild--de Sitter spacetime using three different methods. The first one is proposed in Refs.~\cite{Gregory:2013hja,Burda:2015isa, Burda:2015yfa} and is the full calculation taking into account the backreaction to the metric. The other two are calculations in certain limits: a flat spacetime limit and a fixed-background limit. 
The bubble nucleation rate in 
these latter two methods 
can be decomposed into two factors: 
a quantum tunneling rate from a finite energy $E$ 
and a probability of producing states with an energy $E$. 
The latter factor is just the Boltzmann factor in a finite-temperature system. 
Comparing these results with that of the full calculation used in the literature, 
we clarified the physical meaning of the enhancement factor due to the existence of BH. 
Namely, the enhancement factor can be interpreted as 
a probability of producing states with an energy $\Delta M$, 
where $\Delta M$ is the difference of BH mass before and after the transition. 
This makes it clear that all the other states, such as plasma, are also generated through the same mechanism, and calls for finite-density corrections to the tunneling rate, which tend to stabilize the false vacuum.
We showed also that the probability is just equal to the Boltzmann factor for the energy $E = \Delta M$. 
This means that the results of the latter two calculations coincide with that of the former only when we consider an activation due to the finite temperature and the temperature should be identified with the Hawking temperature associated with the BH horizon. This implies that, in the former calculation, the finite-temperature effect is implicitly taken into account. The enhancement of the bubble nucleation rate should be 
related to the Hawking radiation.

We showed also that the periodicity of the bounce solution is not necessarily related to the temperature of the system, but the consistency of those limits indicates that the bounce solutions around a BH coincide with the thermally activated tunneling associated with the BH Hawking temperature. Although the periodicity of {\it bounce solutions} as a function of its energy can be different from the one indicated by the conical singularity of BH horizon, we find that
the dominant process can always be embedded in the Euclidean \textit{spacetime} with a periodicity of the Hawking temperature.

It may be instructive to interpret two methods in the curved spacetime in terms of statistical mechanics. The former, where we fully include the effect of gravity, may correspond to the microcanonical ensemble. The energy is fixed and should be conserved before and after the transition. This is satisfied in the above calculation because the energy of the bubble is equal to the mass difference of BH $\Delta M$. This interpretation is also supported by the fact that the transition is determined by the change of the BH entropy above a certain threshold. The other method, where we neglect the backreaction to the metric and consider a thermal activation, may correspond to the canonical ensemble, where temperature is fixed but the energy of the bubble is not necessarily conserved before and after the transition. In this case, the free energy should be minimized for the dominant transition process. In any case, the result should be the same in a limit where the backreaction can be neglected because these interpretations are irrelevant for the physical results.

\begin{figure}[t]
\centering 
\includegraphics[width=.95\textwidth]{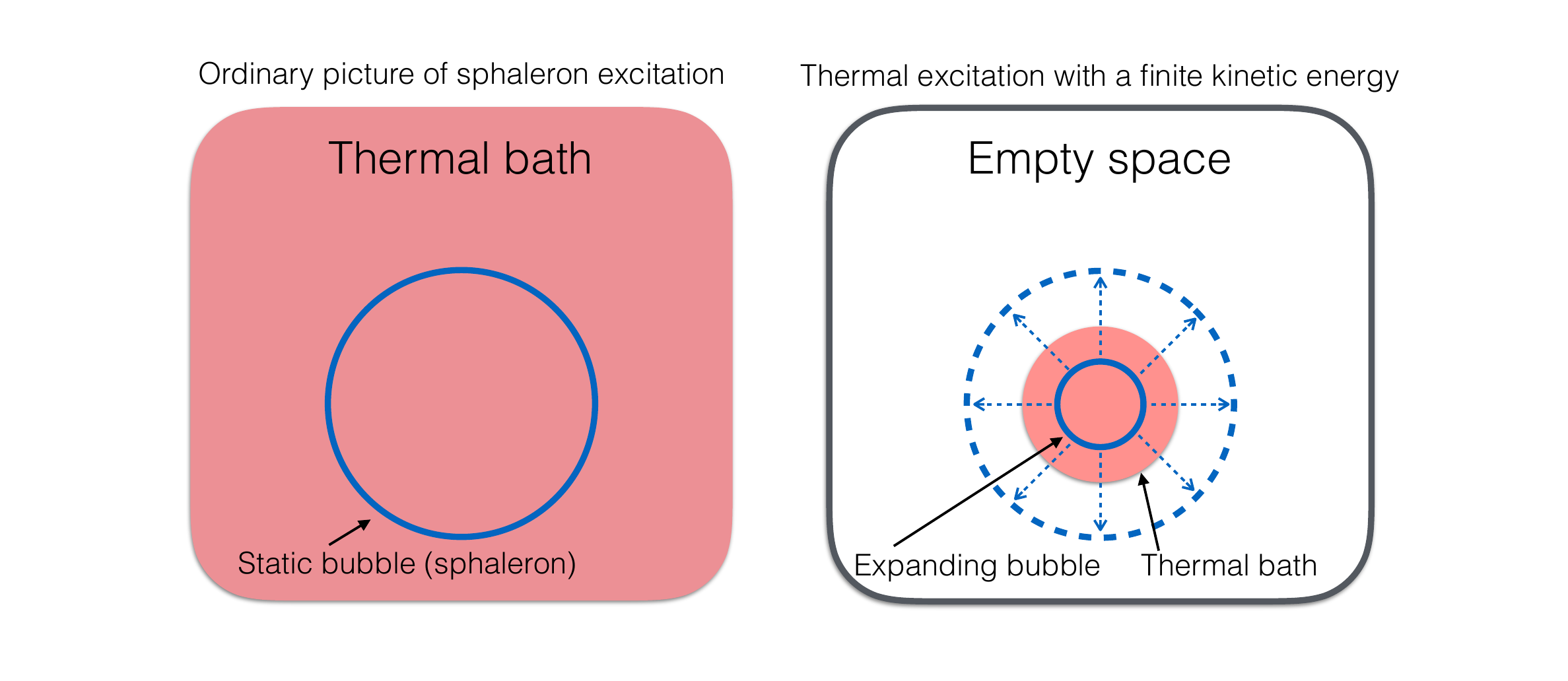} 
\caption{\small
 Schematic figures that describe the excitation of a sphaleron bubble 
 in the thermal plasma with an infinite volume (left panel) 
 and the one in the thermal plasma with a finite volume whose size is smaller than the sphaleron solution (right panel). 
 The process that the bubble nucleation at the BH radius by the Hawking radiation 
 corresponds to is described in the right panel. 
}
  \label{fig:schematic}
\end{figure}

According to our results and the above discussion, we conclude that the bubble nucleation around a BH is associated with a thermal effect of the Hawking temperature,
and the enhancement factor is nothing but a probability of generating bubbles with a finite energy.
The point here is that all the states other than bubbles should also be generated 
since it is hard to imagine a mechanism which only activates bubbles though we have many other degrees of freedom in quantum field theory.
There are two possible interpretations of the probability function. The enhancement may be due to a thermal plasma which fills a whole space outside of the BH as shown in the left panel of Fig.~\ref{fig:schematic}. On the other hand, it may be possible that the bubble nucleation occurs at the BH horizon with a nonzero kinetic energy and it expands to a critical bubble (see the right panel in Fig.~\ref{fig:schematic}). If the former interpretation is correct, the obtained bounce solutions may not correspond to the realistic case, where a BH resides in an almost empty space, but rather to the case where a BH is surrounded by the thermal plasma of the BH Hawking temperature. 
This has been pointed out in Refs.~\cite{Brown:2007sd, HenryTye:2008xu, Masoumi:2012yy} in de Sitter background. 
Moreover, the existence of such thermal plasma may reduce the probability of bubble nucleations. 
Its effect typically makes the transition difficult because the scalar field prefers a symmetric point in the field space due to the thermal mass. As a result, the bubble nucleation rate may not be so drastically enhanced even around a small BH. This viewpoint and question are also discussed in Ref.~\cite{Gorbunov:2017fhq} (see also Refs.~\cite{Tetradis:2016vqb, Canko:2017ebb}). On the other hand, if the latter interpretation is correct, the Universe is not necessarily filled with the thermal plasma and the bounce solutions can be applied to the realistic case. Even in this case, we need to take into account the finite-density corrections to the effective potential because there exists a flux of Hawking radiation emitted from the BH as done in Ref.~\cite{Cheung:2013sxa} (see \textit{e.g.}, Refs.~\cite{Candelas:1980zt, Moss:1984zf, MacGibbon:2007yq, Giddings:2015uzr, Emelyanov:2016tws} for related work). Notice that, however, this interpretation leaves an open question whether or not a bubble can be excited with a nonzero kinetic energy by a finite-volume thermal bath (or at the BH horizon) whose size is much smaller than the critical bubble, and expands to a critical bubble. 
Nevertheless, we postpone the conclusion in this paper; rather we explain how to include the finite-density corrections of Hawking radiation. Detailed studies on this aspect will be presented elsewhere~\cite{future-work}.

Thermal effects originate from the ambiguity of the vacuum state at the false vacuum. When we perform the Wick rotation and calculate the Euclidean action, we use a specific metric where the components of the metric do not depend on the (imaginary) time variable. However, it is well known that the vacuum state has an observer dependence~\cite{Unruh:1976db, Gibbons:1976ue, Gibbons:1977mu, Bunch:1978yq}. For example, a vacuum state defined by a freely falling observer around a BH is a thermally excited state for a static observer. This ambiguity of the vacuum state may add an implicit assumption on the initial state when we calculate a Euclidean action in general relativity. Regarding the observer dependence, we have also obtained a consistent result in the case of bubble nucleations in a de Sitter Universe which supports this viewpoint. Our consideration clarifies that we have to take particular care of the initial vacuum state in the Wick-rotated Euclidean spacetime to calculate the bubble nucleation rate around a BH. This is also supported by the fact that a no-boundary wave function of BH in the Euclidean geometry coincides with the Hartle-Hawking vacuum state~\cite{Hartle:1976tp, Barvinsky:1994jca}.

Strictly speaking, a BH surrounded by an infinite thermal plasma of the BH Hawking temperature
in a de Sitter or flat spacetime is thermodynamically unstable.
This is because a larger/smaller BH gets fat/light by accreting/emitting particles from/to the thermal plasma,
and thus the system is unstable under the perturbation.
Although we naively expect that this effect could be neglected
if one restricts the validity of the calculation to the case where
the lifetime of our vacuum is much shorter than the evaporation time scale of the BH,
a more rigorous way to test the procedure in Refs.~\cite{Gregory:2013hja,Burda:2015isa,Burda:2015yfa} may be 
to study the anti-de Sitter Schwarzschild metric as done recently in Ref.~\cite{Chen:2017suz}.

\section*{Acknowledgments}
\small
We would like to thank Jun'ichi Yokoyama for discussion at an early stage of this work.
This work is supported by Grant-in-Aid for Scientific Research 
from the Ministry of Education, Science, Sports, and Culture
(MEXT), Japan, 
World Premier International Research Center Initiative
(WPI Initiative), MEXT, Japan, 
and the JSPS Research Fellowships for Young Scientists (M.Y. and K.M.).

\small
\bibliography{reference}
  
\end{document}